\documentclass[twocolumn,secnumarabic,amssymb,nobibnotes,aps,prd,superscriptaddress,nofootinbib]{revtex4-2}

\usepackage[utf8]{inputenc}
\usepackage{graphicx,amssymb,amsmath,amsthm,amsfonts,epsfig,epsf}
\usepackage[usenames,dvipsnames,svgnames,table]{xcolor}
\usepackage{epstopdf}
\definecolor{darkred}{rgb}{0.5,0,0}
\usepackage{bm}
\usepackage{dcolumn}
\usepackage{latexsym}
\usepackage{rotating}
\usepackage{longtable}

\usepackage{enumerate}
\usepackage{tensor,multirow}
\usepackage{url}
\usepackage[linktocpage]{hyperref}

\hypersetup{
	colorlinks,
	linkcolor={red!60!black},
	citecolor={green!50!black},
	urlcolor={blue!80!black}
}

\def\nn{\nonumber}

\def\be{\begin{equation}}
\def\ee{\end{equation}}
\newcommand{\beq}{\begin{eqnarray}}
\newcommand{\eeq}{\end{eqnarray}}
\def\ba{\begin{align}}
\def\ea{\end{align}}

\newcommand{\acs}{\textsc}
\newcommand{\eg}{e.g.}
\newcommand{\ie}{i.e.}

\newcommand{\bracket}[2]{\ensuremath{\left\langle #1 \middle| #2 \right\rangle}}

\begin{document}
	
\title{Dynamical friction of black holes in ultralight dark matter}

\author{Rodrigo Vicente}
\email{rodrigoluisvicente@gmail.com}	
\affiliation{Institut de Fisica d’Altes Energies (IFAE), The Barcelona Institute of Science and Technology, Campus UAB, 08193 Bellaterra (Barcelona), Spain}
\author{Vitor Cardoso}%
\email{vitor.cardoso@tecnico.ulisboa.pt}
\affiliation{Centro de Astrof\'{\i}sica e Gravita\c c\~ao - CENTRA, Departamento de F\'{\i}sica, Instituto
Superior T\'ecnico - IST, Universidade de Lisboa - UL, Avenida Rovisco Pais 1, 1049-001 Lisboa, Portugal}
\affiliation{Niels Bohr International Academy, Niels Bohr Institute, Blegdamsvej 17, 2100 Copenhagen, Denmark}
\date{\today}%

\begin{abstract}
In this work we derive simple closed-form expressions for the dynamical friction acting on black holes moving through ultralight (scalar field) dark matter, covering both nonrelativistic and relativistic black hole speeds. Our derivation is based on long known scattering amplitudes in black hole spacetimes, it includes the effect of black hole spin and can be easily extended to vector and tensor light fields. Our results cover and complement recent numerical and previous nonrelativistic treatments of dynamical friction in ultralight dark matter.
\end{abstract}

\maketitle

\section{Introduction}\label{sec:intro}
The search for new interactions has been a vibrant field for decades, the importance of which is hard to overemphasize. 
New axionic degrees of freedom, for example, have been predicted to arise in extensions of the Standard Model~\cite{Peccei:1977hh,PhysRevLett.40.279,PhysRevLett.40.223,Marsh:2015xka,Freitas:2021cfi}.
In fact, a variety of new scalars could populate the Universe~\cite{Arvanitaki:2009fg}. If such new degrees of freedom are ultralight, they would also be a natural component of dark matter (\acs{dm})~\cite{Preskill:1982cy,Abbott:1982af,Dine:1982ah}.
These are often referred to as fuzzy \acs{dm} models, and require ultralight bosonic fields (we refer the reader to Refs.~\cite{Robles:2012uy,Hui:2016ltb,Bar:2019bqz,Bar:2018acw,Desjacques:2019zhf,Davoudiasl:2019nlo,Annulli:2020lyc,Ferreira:2020fam,Hui:2021tkt}). Such extensions are not restricted to scalars or axions: models of minicharged dark matter predict the existence of new fermions which possess a fractional electric charge or are charged under a hidden~$U(1)$ symmetry~\cite{DeRujula:1989fe,Perl:1997nd,Holdom:1985ag,Sigurdson:2004zp,Davidson:2000hf,McDermott:2010pa}. These minicharged particles are a viable candidate for cold \acs{dm} and their properties have been constrained by several cosmological observations and direct-detection experiments~\cite{Davidson:2000hf,Dubovsky:2003yn,Gies:2006ca,Gies:2006hv,Perl:2009zz,Burrage:2009yz,Ahlers:2009kh,Haas:2014dda,Kadota:2016tqq,Gondolo:2016mrz}. In some other models, dark fermions do not possess (fractional) electric charge but interact among each other only through the exchange of dark photons, the latter being the mediators of a long-range gauge interaction with no coupling to Standard Model particles~\cite{Ackerman:mha}.

With the above as motivation, a substantial amount of work has been dedicated to understand the physics of extended scalar structures. For example, the structure of composite stars containing boson stars in their interior is important to understand how dark matter could pile up and change the composition of neutron stars~\cite{Brito:2015yga,Brito:2015yfh,DiGiovanni:2021ejn}. For compact configurations, it is important to understand possible gravitational-wave signatures upon coalescence with black holes or other boson stars~\cite{Cardoso:2016oxy,Palenzuela:2017kcg}. When the configuration is dilute however, numerical simulations become extremely challenging or impossible to perform, as the physical effects of interest act on much longer timescales. Of particular importance is dynamical friction (\acs{df}) and energy loss via scalar emission, which control the motion of objects moving within extended scalar structures~\cite{Hui:2016ltb,Lancaster:2019mde,Annulli:2020lyc,Traykova:2021dua,Chowdhury:2021zik,Wang:2021udl,Baumann:2021fkf}.

To overcome the different and disparate length scales in any astrophysical scenario, recent numerical simulations modeled the spacetime as a fixed Schwarzschild black hole (\acs{bh}) geometry, moving at constant velocity through a scalar field environment of infinite extent, and extracted numerically the \acs{df}~\cite{Traykova:2021dua}. Here, we show that in this setup the \acs{df} can be obtained analytically. We derive simple expressions, valid both for nonrelativistic and relativistic \acs{bh} speeds, from scattering amplitudes in \acs{bh} spacetimes. We focus on stationary regimes and extend the Newtonian expressions in Refs.~\cite{Hui:2016ltb,Lancaster:2019mde}. Our results complement the recent numerical work of Ref.~\cite{Traykova:2021dua}.  

In this work we follow the conventions of Ref.~\cite{Wald:1984rg}; in particular, we adopt the mostly positive metric signature and use geometrized units ($c=G=1$).

\section{Framework}\label{sec:fw}

Ultralight bosons produced through the misalignment mechanism are described by a coherent state~\cite{Davidson:2014hfa,Hui:2021tkt}, for which the relative quantum field fluctuations are suppressed with~$1/\sqrt{N}$, where~$N$ is the (average) occupancy number of the state~\cite{Glauber:1963tx}. From observations we know that the local \acs{dm} density in the Solar System's neighborhood is~$\sim1\,\text{GeV}/\text{cm}^3$~\cite{Bovy:2012tw,Sivertsson:2017rkp,McKee_2015}. On the other hand, virialized ultralight particles in the Galaxy have a de Broglie wavelength~$\lambda_\text{dB} \sim \text{kpc}\left(\frac{10^{-22}\text{eV}}{m_S}\right)\left(\frac{250\, \text{km/s}}{v}\right)$. So, if these ultralight bosons are all the~\acs{dm}, the typical occupancy number is~$N\sim 10^{96} \left(\frac{10^{-22}\text{eV}}{m_S}\right)^4 \left(\frac{250\, \text{km/s}}{v}\right)^3$. Thus, this system can be completely described in terms of a classical field~\cite{Hui:2021tkt}. 

In this work we model the scalar particles through a massive complex scalar field described by the action 
\begin{equation}\label{theory_action}
S= \int d^4x \sqrt{-g}\left(\frac{R}{8 \pi}- \Phi^*_ {\;\,, \alpha} \Phi^{\,,\alpha}-\mu^2|\Phi|^2\right),
\end{equation}
where~$m_S=\hbar \mu$ is the mass of the scalar. Therefore, the scalar field satisfies the Klein-Gordon (\acs{kg}) equation
\begin{equation} \label{KG}
\Box \Phi= \mu^2\Phi,
\end{equation}
and the spacetime metric satisfies the Einstein equations
\begin{equation}
G^{\alpha \beta}=8 \pi \,T^{\alpha \beta} \label{Einstein},
\end{equation}
where~$G^{\alpha \beta}\equiv R^{\alpha \beta} -\frac{1}{2} Rg^{\alpha \beta}$ is the Einstein tensor, and the scalar's energy-momentum tensor is
\begin{equation}
T^{\alpha \beta}= \nabla^{(\alpha}\Phi^* \nabla^{\beta)} \Phi-\frac{1}{2}g^{\alpha \beta} \left(\Phi^*_ {\;\,; \delta} \Phi^{\,;\delta}+\mu^2|\Phi|^2\right). \label{scalarEMT}
\end{equation}

For most situations of interest, the scalar is not very dense and can be studied in a \emph{fixed} spacetime geometry -- the so-called \emph{test field} approximation. So, let us consider a fixed background metric describing a stationary spinning (Kerr)~\acs{bh} with line element, in Boyer-Lindquist coordinates,
\begin{eqnarray}\label{metric}
ds^2=&&-\frac{\Delta}{\rho_r^{2}}(dt-a \sin^2 \theta\, d\varphi)^2+\frac{\rho_r^2}{\Delta}dr^2 \nonumber \\
&&+\rho_r^2d\theta^2+\frac{\sin^2 \theta}{\rho_r^2}\Big(a dt - (r^2+a^2)d\varphi\Big)^2,
\end{eqnarray}
where~$\rho_r^2=r^2+a^2\cos^2 \theta$ and~$\Delta=r^2+a^2-2Mr$, with $0\leq a\leq M$. Here,~$M$ is the \acs{bh} mass and~$J=M a$ its angular momentum (pointing along~$\theta=0$).
%
\section{Scalar field scattering off a black hole at rest}\label{sec:BHrest}

A \acs{bh} moving through an infinite homogeneous scalar field medium is equivalent (by applying a Lorentz boost to the \acs{bh} frame) to a plane wave scattering off a \acs{bh} at rest. So, we start by considering the classical problem of a monochromatic plane wave scattering off a Kerr~\acs{bh} in its proper frame. This scattering problem was studied previously in, \eg,  Refs.~\cite{Matzner:1968,Starobinski:1973,Unruh:1976fm,Sanchez:1977si,MTB,Glampedakis:2001cx,Macedo:2013afa,Leite:2016hws}.

Let us consider the multipolar decomposition of a monochromatic scalar field of frequency $\omega$,
\begin{equation} \label{scalar_decomp}
\Phi=\sum_{\ell,m}e^{-i (\omega t-m \varphi)}  \text{Ps}_\ell^m(\cos \theta, \gamma^2)\mathcal{R}_{\ell}^m(r)\,,
\end{equation}
where $\text{Ps}_{\ell}^m(\cos \theta,\gamma^2)$ are (oblate) angular spheroidal wave functions of the first kind satisfying the ordinary differential equation (\acs{ode})~\cite{NIST:DLMF}
\begin{equation} \label{spheroidal}
	\frac{1}{\sin \theta}\frac{d}{d\theta}\Big[\sin \theta \frac{d\text{Ps}}{d \theta} \Big]+\Big[\lambda_{\ell}^m+\gamma^2 \sin^2\theta-\frac{m^2}{\sin^2 \theta}\Big]\text{Ps}=0,
\end{equation}
with regular conditions at $\theta=\{0,\pi\}$, where
\begin{equation}
\gamma\equiv i k_\infty a\,,\qquad k_\infty\equiv \sqrt{\omega^2-\mu^2}\,.
\end{equation}
The eigenvalues~$\lambda_l^m$ ($\ell \geq |m|$) are not known in analytic form; asymptotically ($\ell\to \infty$) they are~$\lambda_\ell^m=\ell(\ell+1)+\tfrac{1}{2}(k_\infty a)^2+\mathcal{O}(\ell^{-2})$~\cite{Berti:2005gp}.
The above decomposition reduces the \acs{kg} equation to a radial \acs{ode} for the functions~$\mathcal{R}_l^m$~\cite{Starobinski:1973},
\begin{eqnarray}
&&\Delta \frac{d}{d r}\left[\Delta\frac{d \mathcal{R}}{d r}\right]+\Big[\omega^2(r^2+a^2)^2-4 a M m \omega r\nonumber \\
&&+(m a)^2 -(\lambda+\mu^2(r^2+a^2)) \Delta\Big]\mathcal{R}=0. \label{KG_rad_alt}
\end{eqnarray}
Performing the change of variables
\begin{equation}
\frac{dr}{d \chi}=\frac{\Delta}{r^2+a^2}, \quad-\infty<\chi<+\infty,
\end{equation}
\begin{equation}
\mathcal{R}=\frac{f}{\sqrt{r^2+a^2}}\,,
\end{equation}
we can rewrite the radial Eq.~\eqref{KG_rad_alt} in the form of a (time-independent) Schrödinger-like equation
\begin{subequations}\label{KG_rad}
\begin{equation} 
\frac{d^2f}{d\chi^2}+k^2(\chi)f=0, \\
\end{equation}
with
\begin{eqnarray}
k^2[\chi(r)]&=&\left(\omega-\frac{m a}{r^2+a^2} \right)^2-\frac{\Delta}{(r^2+a^2)^2} \bigg\{\lambda+\mu^2(r^2+a^2) \nonumber \\
&&-2am \omega +\sqrt{r^2+a^2} \frac{d}{d r}\left[\frac{r \Delta}{(r^2+a^2)^{3/2}}\right]\bigg\},
\end{eqnarray} 
\end{subequations}

The regular solutions to the above equation satisfy the boundary condition~\cite{Unruh:1976fm}
\begin{equation} \label{scalar_infin}
f(\chi \to +\infty)\simeq I e^{- i \left[k_\infty r-\eta \log (2 k_\infty r) \right]}+ R e^{ i \left[k_\infty r-\eta \log (2 k_\infty r)\right] }\,,
\end{equation}
at spatial infinity, and~\cite{Starobinski:1973}
\begin{equation} \label{scalar_horiz}
f(\chi \to -\infty)\simeq T e^{-i (\omega-m \Omega_\text{h}) \chi}\,,
\end{equation}
at the~\acs{bh} event horizon~$r_\text{h}=M+\sqrt{M^2-a^2}$ (the largest real root of~$\Delta$), where~$\Omega_\text{h}\equiv a/(r_\text{h}^2+a^2)$ is the angular velocity of the \acs{bh} event horizon. %
All the above amplitudes are also functions of the angular numbers, $R=R_{\ell}^m, I=I_{\ell}^m$ and~$T=T_{\ell}^m$. We occasionally omit such dependence whenever it is obvious.
Above we defined the useful parameter
\begin{equation}\label{particleness}
\eta\equiv-M\left(\frac{\omega^2+k_\infty^2}{k_\infty}\right).
\end{equation}
Its absolute value is the ratio of the characteristic (gravitational) deflection radius~$\sim M(\omega^2+k_\infty^2)/k_\infty^2$ (as can be read from Eq.~\eqref{radialEOMII}) to the de Broglie wavelength~$1/k_\infty$. For~$\eta^2\gg1$ the scalar field behaves as a beam of classical particles\footnote{More precisely, as we shall see only the modes~$\ell \gg1$ describe classical particles with impact parameter~$b=k_\infty/\sqrt{\ell(\ell+1)}$.} (\emph{particle} limit), whereas for~$\eta^2\ll1$ the wave effects are at their strongest (\emph{wave} limit).

Note that we must consider only frequencies~$\omega>\mu$, which can arrive to spatial infinity and, so, that allow us to define a scattering problem (alternatively, this is enforced by the Lorentz boost from the scalar's proper frame to the \acs{bh} frame).
The ratios~$R/I$ and~$T/I$ are fixed by Eqs.~\eqref{KG_rad} (or, equivalently, by~\eqref{KG_rad_alt}) and can always be obtained numerically, \eg, by solving Eq.~\eqref{KG_rad_alt} with boundary condition~\eqref{scalar_horiz} (where one can put~$T=1$, using the linearity of the \acs{ode}) and comparing the numerical solution with~\eqref{scalar_infin}.
It is easy to show -- through the conservation of the Wronskian -- that the amplitudes satisfy the relation
\begin{equation} \label{wronsk}
\left|\frac{T}{I}\right|^2=\frac{k_\infty}{\omega-m \Omega_\text{h}}\left(1-\left|\frac{R}{I}\right|^2\right).
\end{equation}

A monochromatic plane wave of frequency~$\omega$ and wave vector~$\bm{k
}_\infty=k_\infty\bm{\xi}$, where~$\bm{\xi}=-\cos\beta {\partial}_ z+\sin \beta {\partial}_y$,\footnote{Without loss of generality, due to the axial symmetry with respect to the \acs{bh}'s rotation axis~${\partial}_z$.} deformed by a long-range gravitational potential~$\eta/r$ can be written in the form~\cite{Matzner:1968,Glampedakis:2001cx}
\begin{widetext}
\begin{eqnarray} 
&&e^{-i \left[k_\infty r-\eta \log (2 k_\infty r)\right] \left(\cos\beta \cos \theta+ \sin\beta \sin \theta \sin \varphi\right)} \simeq \nonumber \\
&&\simeq\bigg(\frac{e^{-i \left[k_\infty r-\eta \log (2 k_\infty r) \right]}}{k_\infty r}\bigg) \sum_{\ell, m}(-i)^{m+1}\frac{2\ell+1}{2}\frac{(\ell-m)!}{(\ell+m)!}  \text{Ps}_\ell^m(\cos\beta,\gamma^2)\text{Ps}_\ell^m(\cos \theta,\gamma^2) +\, \textrm{(outgoing wave)}.
\end{eqnarray}
\end{widetext}
So, choosing the incident amplitude
\begin{equation} \label{incident_ampl}
I=\frac{2\ell+1}{2} \sqrt{\frac{\hbar n}{\mu}}\frac{(\ell-m)!}{(\ell+m)!} \frac{(-i)^{m+1} \text{Ps}_\ell^m(\cos\beta,\gamma^2)}{k_\infty} ,
\end{equation}
the solution~\eqref{scalar_decomp} describes a beam of scalar particles with proper number density~$n$ and momentum~$\hbar\bm{k}_\infty$ scattering off a Kerr \acs{bh}. Note that we need to include the amplitude~$\sqrt{\hbar n/\mu}$, so that the energy density current of the plane wave is~$\lim_{r \to \infty}(-T_{t\alpha} \xi^\alpha)=(n k_\infty/\mu) (\hbar \omega)$, \ie, the product of the number density current~$n k_\infty/\mu$ and the energy of each particle~$\hbar \omega$.

\vskip 1cm
\subsection{Energy absorption}
The energy of the scalar field contained in a spacelike hypersurface~$\mathcal{S}_{t'}\equiv\{t=t'\}$ extending from the horizon to infinity is
\begin{equation}
E(t')=\int_{\mathcal{S}_{t'}}dV_3\, T^{\alpha t} N_\alpha\,,
\end{equation}
where~$N_\beta=-\delta_\beta^t/\sqrt{-g^ {tt}}$ is the unit normal covector and~$dV_3$ is the volume form induced in the hypersurface. Because we are considering the scattering of monochromatic waves~$\Phi\propto e^{-i\omega t}$, which results in a time-invariant energy-momentum tensor~$T^{\alpha \beta}$, and since the background metric is stationary, one has
\begin{equation}
\frac{d E}{d t'}=\int_{\mathcal{S}_{t'}}\mathcal{L}_{\partial_ t} \left(dV_3\, T^{\alpha t} N_\alpha\right)=0\,,
\end{equation}
where~$\mathcal{L}_{\partial_t}(\cdot)$ is the Lie derivative with respect to~$(\partial_t)^\alpha$. Then, by applying the divergence theorem it follows that~$\dot{E}_\text{BH}$, the energy crossing the event horizon per unit time~$t$ (the proper time of a stationary observer at infinity), is equal to the flux through a 2-sphere at spatial infinity 
\begin{equation}
\dot{E}_\text{BH}=\int_{r \to \infty}  d\Omega_ 2 \,r^2 T_{rt},
\end{equation}
with the element of area~$d\Omega_ 2=\sin\theta\, d\theta d\varphi$.

Plugging the decomposition~\eqref{scalar_decomp} with the asymptotic solution~\eqref{scalar_infin} in the scalar field's energy-momentum tensor~\eqref{scalarEMT} and using the orthogonality relations of the oblate angular spheroidal wave functions~\cite{NIST:DLMF}
\begin{equation}
	\int_0^\pi d\theta \sin \theta\, \text{Ps}_\ell^m \text{Ps}_{\ell'}^m=\frac{2}{2\ell+1} \frac{(\ell+m)!}{(\ell-m)!} \,\delta_{\ell,\ell'}
\end{equation}
it is straightforward to show that
\begin{equation}
\dot{E}_\text{BH}=\omega k_\infty \sum_{\ell,m} \frac{4\pi }{2\ell+1} \frac{(\ell+m)!}{(\ell-m)!}\left(|I|^2-|R|^2\right).
\end{equation}
For the case of an incident monochromatic beam described by the amplitude~\eqref{incident_ampl}, the last expression becomes
\begin{equation} \label{dotE}
\dot{E}_\text{BH}= \frac{\pi \hbar\omega n}{\mu k_\infty}\sum_{\ell,m}  (2\ell+1)\frac{(\ell-m)!}{(\ell+m)!}(\text{Ps}_\ell^m)^2\bigg(1-\bigg|\frac{R}{I}\bigg|^2\bigg).
\end{equation}

As a consistency check: note that in a flat spacetime (\ie, $M=0$) the plane wave propagates freely and it is easy to show that~$R/I=(-1)^{\ell+1}$, which implies~$\dot{E}_\text{BH}=0$ as expected (since there is no~\acs{bh} at all); moreover, in the case of a static~\acs{bh}, due to spherical symmetry, one can choose~$\beta=0$ without loss of generality, which results in~$\text{Ps}_\ell^m\propto\delta_{m,0}$ and the above expression reduces then to the one found in, \eg,  Refs.~\cite{Unruh:1976fm,Vicente:2021mqd}. 

To obtain the \acs{bh} absorption cross section we just need to take the ratio of the energy absorbed by the~\acs{bh} per unit of time~$\dot{E}_\text{BH}$ to the energy density current of the incident plane wave~$(n k_\infty/\mu) (\hbar \omega)$,
\begin{equation}\label{cross_abs}
\sigma_{\textrm{abs}}=\frac{\pi}{k_\infty^2}\sum_{\ell,m}  (2\ell+1)\frac{(\ell-m)!}{(\ell+m)!}(\text{Ps}_\ell^m)^2\bigg(1-\bigg|\frac{R}{I}\bigg|^2\bigg).
\end{equation}

\vskip 1cm
\subsubsection{Low-frequency limit~($\omega M \ll 1$)} 
For sufficiently low frequencies~$\omega M \ll 1$ one can use matched asymptotic expansions to obtain analytical expressions for the scattering amplitudes (worked out in Appendix~\ref{app:LF}). So, using the reflection amplitude~\eqref{R_lower}, at leading order in~$\omega M$, the energy absorbed by the~\acs{bh} is
\begin{equation}\label{accretion}
\dot{E}_\text{BH} \simeq \frac{\hbar \omega^2 n  A_\text{h}}{\mu} \frac{e^{-\pi \eta}\pi \eta}{\sinh(\pi \eta)},
\end{equation}
and the \acs{bh}'s absorption cross section is
\begin{equation}\label{crosssection}
\frac{\sigma_{\textrm{abs}}}{A_\text{h}}\simeq  \frac{\omega }{k_\infty} \frac{e^{-\pi \eta}\pi \eta}{\sinh(\pi \eta)},
\end{equation}
where~$A_\text{h}=4 \pi(r_\text{h}^2+a^2)$ is the event horizon area.
To get this result, we used the fact that at leading order in~$\omega M$ only the $l=0$ mode contributes to both~$\dot{E}_\text{BH}$ and~$\sigma_{\textrm{abs}}$, and the spheroidal wave functions become then~$\text{Ps}_0^0(\cos \beta,\gamma^2)\simeq 1$. We also made use of the property~$|\Gamma(1+i \eta)|^2=\pi \eta/\sinh(\pi \eta)$. The particle and wave limits of the factor~$e^{-\pi \eta} \pi \eta/ \sinh(\pi \eta)$ are shown in Table~\ref{tb:crosssection}.

\begin{table}[h]
	\centering
	\begin{tabular}{c|c}
		\hline
		\hline
		 Particle ($\eta^2\gg1$) & Wave ($\eta^2\ll1$)   \\ 
		\hline
		
		  $2\pi(-\eta)$                                & $1$ \\

		\hline
		\hline
	\end{tabular} 
	\caption{The factor~$e^{-\pi \eta}\pi \eta/\sinh(\pi \eta)$ in the particle and wave limits (c.f., Eqs.~\eqref{accretion} and~\eqref{crosssection}).}
	\label{tb:crosssection}
\end{table}

Note that in the limit~$\omega \gg \mu$ we recover the famous result~$\sigma_{\textrm{abs}} \simeq A_\text{h}$ derived for massless scalar fields in the low-frequency~$\omega M \ll 1$ limit\footnote{The condition~$\omega \gg \mu$ is in the wave regime, since we are in the low-frequency limit~$\omega M\ll1$ (c.f.,~\eqref{particleness}).}~\cite{Starobinski:1973,Das:1996we}. For Schwarzschild \acs{bh}s ($a=0$) the expressions of this section reduce to the ones obtained decades ago by Unruh~\cite{Unruh:1976fm}. At leading order, the spin dependence of the \acs{bh}'s absorption cross section is encoded solely in the event horizon area.

We verified that the analytical approximation obtained in Appendix~\ref{app:LF} through matched asymptotic expansions describes perfectly (with an error~$\leq 1 \%$) the numerical values of~$1-|R_\ell/I_\ell|^2$ for $\omega M/(\ell+1) \leq 0.01$. Unfortunately, this approximation rapidly breaks down for larger frequencies; \eg, for~$\omega M/(\ell+1) \sim 0.05$ our approximation underestimates in~$\sim 15\%$ the true numerical value of~$1-|R_\ell/I_\ell|^2$.  

\vskip 1cm
\subsubsection{High-frequency limit~($\omega M \gg 1$)}
For high frequencies~$\omega M \gg 1$ one can use the Wentzel-Kramers-Brillouin (\acs{wkb}) approximation to obtain analytical expressions for the scattering amplitudes (as done in Appendix~\ref{app:LF}). Using the approximation~\eqref{ratio_high_freq} for the reflection amplitude we can directly evaluate Eqs.~\eqref{dotE} and~\eqref{cross_abs}; this approximation assumes additionally that~$\omega \gg \mu$ (which is necessarily true for scalars with~$\mu M \leq 1$). For a general incident angle~$\beta$ and spin parameter~$a$ it is difficult to proceed analytically and one is forced to evaluate these expressions numerically. However, if we restrict to the particular case of small incident angles one can still do a semianalytical treatment; so, we will focus on this particular case.

\paragraph*{\textsc{small incident angles ($\beta \omega M \ll 1$)}}
From this assumption one has~$\text{Ps}_\ell^m(\cos \beta,\gamma^2)\simeq \delta_{m 0} \text{Ps}_\ell^0(0,\gamma^2)$ and that most of the contribution to the summation in~$\ell$ comes from large~$\ell\gtrsim \omega M$, in which case $\text{Ps}_\ell^0(0,\gamma^2)\simeq 1$.
Approximating the sum by an integral (an excellent approximation at large~$\ell$) we find
\begin{eqnarray} \label{E_quantum_rest2}
	\dot{E}_\text{BH}&\simeq& \frac{2 \pi \hbar n (\omega M)^2}{\mu} \int_0^{\left(\ell/\omega M\right)_\text{cr}} d\Big(\frac{\ell}{\omega M}\Big)\frac{\ell}{\omega M}\nn\\
	&=&\frac{\pi \hbar n(\omega M)^2 \left(\ell/\omega M\right)_\text{cr}^2}{\mu},
\end{eqnarray}
\begin{eqnarray}
	\sigma_{\textrm{abs}}&\simeq& 2\pi M^2\int_0^{\left(\ell/\omega M\right)_\text{cr}} d\Big(\frac{\ell}{\omega M}\Big)\frac{\ell}{\omega M}\nn \\
	&=&\pi M^2 \left(\frac{\ell}{\omega M}\right)_\text{cr}^2. 
\end{eqnarray}
Here, the critical impact parameter~$\left(\ell/\omega M\right)_\text{cr}$ is evaluated at~$\beta=0$ and can be obtained numerically as a function of the dimensionless spin parameter~$\tilde{a}\equiv a/M$ through the procedure described in Appendix~\ref{app:HF}; the result is well fitted by the expression
\begin{equation}
	\left(\frac{\ell}{\omega M}\right)_\text{cr} \simeq 3\sqrt 3-0.28\, \tilde{a}^2-0.087\, \tilde{a}^4,
\end{equation}
which is accurate to~$0.08\%$ in the whole range of~$\tilde{a}$.
For a nonspinning \acs{bh} ($\tilde{a}=0$) we recover the well-known result~$\sigma_{\textrm{abs}}\simeq 27\pi M^2$~\cite{Sanchez:1976}.
We find that the \acs{bh} spin leads to a decrease in the energy absorption; in the particular case of small incident angles, the spin can suppress the absorption by up to~$13.6\%$.

%
%

\vskip 1cm
\subsection{Transfer of momentum}
In the scattering process there will be a transfer of momentum from the scalar field to the~\acs{bh}, and so the latter will feel a \emph{force}. Consider the spatial components of the Arnowitt-Deser-Misner (\acs{adm}) momentum~$P^i$ computed using a 2-sphere with a sufficiently large radius. These components can be decomposed into the sum of curvature and scalar field contributions~$P^i=P_\text{BH}^i+ P_S^i$, where~$P_S^i$ is 
\begin{equation}
P_S^i(t')=\int_{S_{t'}} dV_3 T^{\alpha i} N_\alpha.
\end{equation}
The rate of change of~$P^i$ is
\begin{equation}
\frac{d P^i}{d t'}=-\int_{r\to \infty} d\Omega\, r^2 T^{r i} ,
\end{equation}
and, because we considering a stationary regime, we have
\begin{equation}
\dot{P}_S^i(t')=\int_{\mathcal{S}_{t'}}\mathcal{L}_{\partial_ t} \left(dV_3\, T^{\alpha i} N_\beta\right)=0.
\end{equation}
Thus, the force acting on the~\acs{bh} is
\begin{equation}\label{force_BHframe}
F^i\equiv \dot{P}^i_\text{BH}=\dot{P}^i=-\int_{r\to \infty} d\Omega_ 2\, r^2 T^{r i}.
\end{equation}
Strictly, in the test field approximation one has~$\dot{P}^i_\text{BH}=0$ (at first order in the scalar field) and~$\dot{P}^i\neq 0$ (at second order in the scalar field). This is not inconsistent with the last equation, which holds  at each order in the scalar field. In other words, would we compute the backreaction of the scalar field on the metric, we would obtain a second order correction to~$\dot{P}^i_\text{BH}$ which must be equal to~$\dot{P}^i$ (at the same order). For a more thorough discussion, which also covers the case where the steady state is attained dynamically, see Ref.~\cite{Clough:2021qlv}.

In asymptotic Cartesian coordinates~$(x,y,z)$, defined such that the \acs{bh} angular momentum is~$\bm{J}=J \partial_ z$ and the direction of incidence is~$\bm{\xi}=-\cos\beta {\partial}_ z+\sin \beta {\partial}_y$, we have
\begin{subequations} \label{EMT_comp}
\begin{eqnarray}	
\lim_{r \to \infty}r^2T^{r x} &\simeq& r^2\sin \theta \cos \varphi T_{rr}\nn\\
&=&r^2 \left(\text{P}_1^{-1} e^{i\varphi}-\frac{\text{P}_1^1}{2} e^{-i\varphi}\right)T_{rr} , 
\end{eqnarray}
\begin{eqnarray}
\lim_{r \to \infty}r^2T^{r y} &\simeq& r^2\sin \theta \sin \varphi T_{rr}\nn \\
&=&-i\,r^2 \left(\text{P}_1^{-1} e^{i\varphi}+\frac{\text{P}_1^1}{2} e^{-i\varphi}\right)T_{rr}, 
\end{eqnarray}
\begin{eqnarray}
\lim_{r \to \infty}r^2T^{r z} &\simeq& r^2\cos \theta T_{rr}=r^2 \text{P}_1^0\, T_{rr},\nn \\ 
\end{eqnarray}
\end{subequations}
where the~$\text{P}_\ell^m(\cos \theta)$ are associated Legendre polynomials~\cite{NIST:DLMF}.
To evaluate the integrals~\eqref{force_BHframe} we will make use of the identity
\begin{eqnarray} \label{WignerEckart}
	&&\int d\Omega\, e^{-i(m_1+m_2-m_3)\varphi}\text{P}_{\ell_1}^{m_1}\text{P}_{\ell_2}^{m_2}\text{P}_{\ell_3}^{m_3}=\tfrac{ 4\pi}{2 \ell_3+1} \sqrt{\tfrac{(\ell_3+m_3)!}{(\ell_3-m_3)!}} \nn\\
	&&\times\sqrt{\tfrac{(\ell_1+m_1)!}{(\ell_1-m_1)!} \tfrac{(\ell_2+m_2)!}{(\ell_2-m_2)!}}\bracket{\ell_1\,0\,\ell_2\,0}{\ell_3\,0} \bracket{\ell_1\,m_1\,\ell_2\,m_2}{\ell_3\,m_3}, \nn \\
\end{eqnarray}
where~$\bracket{\ell_1\,m_1\,\ell_2\,m_2}{\ell_3\,m_3}$ are Clebsch-Gordan coefficients; the last identity is a direct consequence of the Wigner-Eckart theorem.
Plugging the decomposition~\eqref{scalar_decomp} with the asymptotic form~\eqref{scalar_infin} in the scalar field's energy-momentum tensor~\eqref{scalarEMT} we find
\begin{eqnarray} \label{Trr}
\lim_{r\to \infty} r^2 T_{rr}&=&k_\infty ^2 \sum_{\ell, m,\ell',m'} \text{Ps}_{\ell'}^{m'}\text{Ps}_\ell^m e^{-i(m'-m) \varphi}\nn \\
&&\times \left(I_{\ell'}^{m'*} I_\ell^m+R_{\ell'}^{m'*} R_\ell^m\right).
\end{eqnarray}

\vskip 1cm
\subsubsection{Low-frequency limit~($\omega M \ll 1$)} 
In this limit it is useful to consider the power expansion of the (oblate) angular spheroidal wave functions in~$\gamma$ (since $|\gamma |\sim \mathcal{O}(k_\infty M)\ll1$)\footnote{There is a global sign mistake in the coefficient of~$\gamma^2$ in Eq.~(4) of~\cite{weisstein}.}~\cite{weisstein}
\begin{subequations}\label{Spherod_power}
	\begin{equation}
		\text{Ps}_\ell^m=\text{P}_\ell^m-\left(\text{a}_\ell^m\text{P}_{\ell+2}^m+\text{b}_\ell^m\text{P}_{\ell-2}^m\right)\gamma^2+ \mathcal{O}(\gamma^3),
	\end{equation}
	\begin{equation}
		\text{a}_\ell^m=\frac{(\ell-m+1)(\ell-m+2)}{2(2\ell +1)(2\ell+3)^2},
	\end{equation}
	\begin{equation}
		\text{b}_\ell^m=-\frac{(\ell+m-1)(\ell+m)}{2(2\ell-1)^2(2\ell+1)}.
	\end{equation}
\end{subequations}
Then, substituting the power expansion~\eqref{Spherod_power} in~\eqref{Trr} and using Eqs.~\eqref{EMT_comp} and~\eqref{WignerEckart} it is straightforward to show
\begin{widetext}
\begin{equation}
	F^x=\tfrac{\pi \hbar n}{\mu}\sum_{\ell, m} \bigg\{\tfrac{(\ell-m+2)!}{(\ell+m)!} \text{P}_\ell^{m}\text{P}_{\ell+1}^{m-1} \Im \left[\Big(\tfrac{R_\ell^{m}}{I_\ell^{m}}\Big)^*\tfrac{R_{\ell+1}^{m-1}}{I_{\ell+1}^{m-1}}\right]+\tfrac{(\ell-m)!}{(\ell+m)!}\text{P}_\ell^{m}\text{P}_{\ell+1}^{m+1}\Im \left[\Big(\tfrac{R_\ell^{m}}{I_\ell^{m}}\Big)^*\tfrac{R_{\ell+1}^{m+1}}{I_{\ell+1}^{m+1}}\right]\bigg\}+\mathcal{O}\left(|\gamma|^2\right),
\end{equation}
\begin{equation}
	F^y=-\tfrac{\pi \hbar n}{\mu}\sum_{\ell, m} \bigg\{\tfrac{(\ell-m+2)!}{(\ell+m)!} \text{P}_\ell^{m}\text{P}_{\ell+1}^{m-1} \Re \left[1+\Big(\tfrac{R_\ell^{m}}{I_\ell^{m}}\Big)^*\tfrac{R_{\ell+1}^{m-1}}{I_{\ell+1}^{m-1}}\right]-\tfrac{(\ell-m)!}{(\ell+m)!}\text{P}_\ell^{m}\text{P}_{\ell+1}^{m+1}\Re \left[1+\Big(\tfrac{R_\ell^{m}}{I_\ell^{m}}\Big)^*\tfrac{R_{\ell+1}^{m+1}}{I_{\ell+1}^{m+1}}\right]\bigg\}+\mathcal{O}\left(|\gamma|^2\right),
\end{equation}
\begin{equation}
	F^z=-\tfrac{2\pi \hbar n}{\mu} \sum_{\ell, m} \tfrac{(\ell-m+1)!}{(\ell+m)!} \text{P}_\ell^m \text{P}_{\ell+1}^m\Re \left[1+\Big(\tfrac{R_\ell^{m}}{I_\ell^{m}}\Big)^*\tfrac{R_{\ell+1}^{m}}{I_{\ell+1}^{m}} \right]+\mathcal{O}\left(|\gamma|^2\right),
\end{equation}
\end{widetext}
where the associated Legendre polynomials in the last expressions are evaluated at~$\cos\beta$. The symbols~$\Re$ and $\Im$ stand, respectively, for the real and imaginary parts of a complex number. Using Eq.~\eqref{R_lower} one can see that at linear order in~$\omega M$ the products of reflection amplitudes in the last expressions are independent of~$m$. Using the identity
\begin{align}
	\sum_m \text{P}_\ell^{m}\left[\frac{(\ell-m+2)!}{(\ell+m)!} \text{P}_{\ell+1}^{m-1}+ \frac{(\ell-m)!}{(\ell+m)!}\text{P}_{\ell+1}^{m+1}\right]=0,
\end{align}
and (by the scattering amplitude~\eqref{R_lower})
\begin{eqnarray}
	\Re \left[1+\bigg(\frac{R_\ell}{I_\ell}\bigg)^*\frac{R_{\ell+1}}{I_{\ell+1}}\right]&\simeq&2\sin^2\left(\frac{\alpha_\ell}{2}\right)\nn\\
	&+&\delta_{\ell 0}\frac{ \omega k_\infty A_\text{h}}{2 \pi} \frac{e^{-\pi \eta}\pi \eta}{\sinh(\pi \eta)},
\end{eqnarray}
with the \emph{deflection angle} 
\begin{eqnarray}
	\alpha_\ell\equiv2\arg(\ell+1+i\eta)=2\arctan\left(\frac{\eta}{\ell+1}\right),
\end{eqnarray}
the force components become (at leading order in~$\omega M$)
\begin{equation}
	F^x\simeq 0,
\end{equation}
\begin{eqnarray}
	F^y&\simeq& \frac{4\pi \hbar n}{\mu}\sum_{\ell, m} \frac{(\ell-m)!}{(\ell+m)!}\text{P}_\ell^{m}\text{P}_{\ell+1}^{m+1} \nn\\
	&&\times\left[\sin^2 \left(\frac{\alpha_\ell}{2}\right)+\delta_{\ell 0}\frac{ \omega k_\infty A_\text{h}}{4 \pi} \frac{e^{-\pi \eta}\pi \eta}{\sinh(\pi \eta)}\right],
\end{eqnarray}
\begin{eqnarray}
	F^z&\simeq& -\frac{4\pi \hbar n}{\mu} \sum_{\ell, m} \frac{(\ell-m+1)!}{(\ell+m)!} \text{P}_\ell^m \text{P}_{\ell+1}^m \nn\\
	&&\times\left[\sin^2 \left(\frac{\alpha_\ell}{2}\right)+\delta_{\ell 0}\frac{ \omega k_\infty A_\text{h}}{4 \pi} \frac{e^{-\pi \eta}\pi \eta}{\sinh(\pi \eta)}\right].
\end{eqnarray}
In the Cartesian frame~$(\partial_{x'},\partial_{y'},\partial_{z'})$ obtained by rotating~$(\partial_{x},\partial_{y},\partial_{z})$ by an angle~$\beta$ around~$\partial_x$ (so that the direction of incidence is~$\bm{\xi}=-\partial_{z'}$) the components of the force acting on the \acs{bh} are~$F^{x'}=F^x$,~$F^{y'}=\cos \beta F^y+ \sin \beta F^z$ and~$F^{z'}=\cos \beta F^z - \sin \beta F^y$. Now using the identities
\begin{equation}
	\sum_ m \text{P}_\ell^m \left[\tfrac{(\ell-m)!}{(\ell+m)!}\text{P}_{\ell+1}^{m+1} \cos \beta+\tfrac{(\ell-m+1)!}{(\ell+m)!}\text{P}_{\ell+1}^m \sin \beta\right]=0,
\end{equation}
\begin{equation}
	\sum_ m \text{P}_\ell^m \left[\tfrac{(\ell-m+1)!}{(\ell+m)!}\text{P}_{\ell+1}^m \cos \beta-\tfrac{(\ell-m)!}{(\ell+m)!}\text{P}_{\ell+1}^{m+1} \sin \beta\right]=\ell+1,
\end{equation}
we obtain
\begin{equation}
	F^{x'}\simeq F^{y'}\simeq 0,
\end{equation}
\begin{eqnarray}\label{force_low}
	F^{z'}&\simeq& -\frac{4 \pi \hbar n}{\mu}\left\{\sum_{\ell \geq 1} \ell \sin^2\left(\frac{\alpha_{\ell-1}}{2}\right)+\frac{ \omega k_\infty A_\text{h}}{4 \pi} \frac{e^{-\pi \eta}\pi \eta}{\sinh(\pi \eta)}\right\} \nn \\
	&=&-\frac{4 \pi \hbar n}{\mu}\left\{\sum_{\ell \geq 1} \frac{\eta^2 \ell}{\eta^2+\ell^2}+\frac{ \omega k_\infty A_\text{h}}{4 \pi} \frac{e^{-\pi \eta}\pi \eta}{\sinh(\pi \eta)}\right\}. \nn\\
\end{eqnarray}
We see that at leading order in~$\omega M$ the force acting on the \acs{bh} does not depend on the~\acs{bh} spin nor on its angle with respect to the direction of incidence, and it is directed along the direction of incidence~$\bm{\xi}$. This can be interpreted as a consequence of the fact that in the limit~$\omega M\ll1$ the force acting on the \acs{bh} is imparted mostly by scalar field probing the weak (gravitational) field, which is not sensitive to~$a$. Actually, the~$\ell=0$ mode probes the strong field and is substantially absorbed by the \acs{bh}, being responsible for the extra term in the above expression, but this contribution is also independent of~$a$ because of its spherical symmetry. On the other hand, it is easily seen that the force diverges logarithmically in~$\ell$ -- which is to be expected due to the long-range~$1/r$ nature of the gravitational potential. So, we proceed by introducing a cutoff~$\ell_\text{max}$, which is associated with the size of the incident beam; the maximum impact parameter is roughly~$b_\text{max}=\sqrt{\ell_\text{max}(\ell_\text{max}+1)}/ k_\infty $ (this cutoff scheme is discussed in more detail in Sec.~\ref{sec:discussion}). 
The truncated sum can be written in terms of the digamma function~$\Psi$~\cite{NIST:DLMF}, after which the force becomes
\begin{eqnarray}\label{force_low1}
	F^{z'}\simeq -\frac{4 \pi \hbar n}{\mu}&&\left\{\eta^2\Re\Big[\Psi(1+\ell_\text{max}+i\eta)-\Psi(1+i\eta)\Big] \right. \nn \\
	&&\left.+\frac{ \omega k_\infty A_\text{h}}{4 \pi} \frac{e^{-\pi \eta}\pi \eta}{\sinh(\pi \eta)}\right\}.
\end{eqnarray}
It is easy to show that in the particle and wave limits the finite sum is excellently approximated by the closed-form expressions in Table~\ref{tb:force}.

\begin{table}[h] 
	\centering
	\begin{tabular}{c|c}
		\hline
		\hline
		Particle ($\eta^2\gg1$) & Wave ($\eta^2\ll1$)   \\ 
		\hline
		
		$(1/2)\log\left(1+k_\infty^2 b_\text{max}^2/\eta^2\right)$                                & $ \log(k_\infty b_\text{max})+\gamma_\text{E}$ \\
		
		\hline
		\hline
	\end{tabular} 
	\caption{The term~$\Re\big[\Psi(1+\ell_\text{max}+i\eta)-\Psi(1+i\eta)\big]$ in the particle and wave limits (c.f., Eq.~\eqref{force_low1}); $\gamma_\textrm{E}= 0.5772 \,...$ is Euler's constant~\cite{NIST:DLMF}.}
	\label{tb:force}
\end{table}

It is worth noting that, in the eikonal limit~$\ell\gg1$,~$\alpha_\ell$ is indeed the deflection angle of a particle scattering off a weak gravitational field with impact parameter~$b\simeq\ell/k_\infty$ and angular momentum~$\hbar \ell$.\footnote{The eikonal limit can be seen as a manifestation of Bohr's correspondence principle. For an interesting discussion about the correspondence between wave and particle scattering see Ref.~\cite{FORD1959259}.} In particular, in the nonrelativistic limit ($\omega\sim \mu$) one finds
\begin{equation}
	\alpha_\ell \simeq-2\arctan\left(\frac{M \mu^2}{b \,k_\infty^2 }\right),
\end{equation}
which is exactly the Newtonian deflection angle; and in the ultrarelativistic limit ($\omega \gg \mu$) one gets
\begin{equation}
	\alpha_\ell \simeq -2\arctan \left(\frac{2 M}{b}\right) \simeq - \frac{4 M}{b},
\end{equation}
which is the deflection angle of light rays obtained by~Einstein using his general theory of relativity~\cite{Einstein1936LENSLIKEAO}. If we compute then the force that would act on a source of such weak gravitational field due to a beam of particles coming with momentum~$\hbar k_\infty$ and impact parameters between~$b$ and $b+\delta b$ and being deflected by an angle~$\alpha_\ell$ we find
\begin{equation}
	\frac{\delta F^{z'}}{k_\infty \delta b}=- \frac{4 \pi \hbar n}{\mu}(k_\infty b) \sin^2 \left(\frac{\alpha_\ell}{2}\right),
\end{equation}
which matches the first term of Eq.~\eqref{force_low} in the eikonal limit; the extra term is due to accretion as explained above. 

We verified that the analytic approximation derived in Appendix~\ref{app:LF} and employed here describes quite well (with an error~$\leq 5\%$) the numerical values of~$\Re \left[1+(R_\ell/I_\ell)^* (R_{\ell+1}/I_{\ell+1})\right]$ for~$\omega M \leq 0.01$. As for the energy absorption, the analytic approximation breaks down for larger frequencies; \eg, for~$\omega M/(l+1)\sim 0.05$ our expression underestimates in~$\sim 20 \%$ the true numerical value of~$\Re \left[1+(R_\ell/I_\ell)^* (R_{\ell+1}/I_{\ell+1})\right]$.

\vskip 1cm
\subsubsection{High-frequency limit~($\omega M \gg 1$)} 
To proceed with a semianalytical treatment we focus again on the case of small incident angles.

\paragraph*{\textsc{small incident angles ($\beta \omega M \ll 1$)}}
In this case we can consider only~$m=0$ modes in the scalar's decomposition~\eqref{scalar_decomp} (due to the approximate axial symmetry) and we note that most of the contribution to the summation in~$\ell$ comes from large~$\ell\gtrsim \omega M$, in which case $\text{Ps}_\ell^0(\cos \theta,\gamma^2)\simeq \text{P}_\ell^0 (\cos \theta)$.
So, using Eqs.~\eqref{EMT_comp},~\eqref{WignerEckart} and~\eqref{Trr} it is straightforward to show
\begin{equation}
	F^x\simeq F^y \simeq 0,
\end{equation}
\begin{equation} \label{force_high}
	F^z \simeq -\frac{2\pi \hbar n}{\mu} \sum_\ell(\ell+1) 	\Re \left[1+\bigg(\frac{R_\ell^0}{I_\ell^0}\bigg)^*\frac{R_{\ell+1}^0}{I_{\ell+1}^0}\right],
\end{equation}
where the reflection amplitudes can be approximated by~\eqref{ratio_high_freq}.
Here the accretion of scalar field gives an important contribution to the force acting on the~\acs{bh}, which is contained in the terms of~\eqref{force_high} with~$\ell<\ell_\text{cr}$, for which~$R/I\simeq 0$. In the high-frequency limit the summation is dominated by the large azimuthal numbers~$\ell\gg1$ and, so, can be approximated by an integral. Thus, the accretion of scalar is responsible for the contribution~$-\pi \hbar n \ell_\text{cr}^2/\mu$ to the force (which, naturally, matches~$\dot{E}_\text{BH}$ in absolute value, since we are considering the ultrarelativistic regime~$\omega\gg\mu$).
The remaining contribution from larger~$\ell$'s can be obtained by using the eikonal approximation~(where~$\ell \simeq \omega b$) and rewriting the summation as
\begin{align}
\sum_{\ell>\ell_\text{cr}} (\ell+1) 
	\Re \left[1+\left(\tfrac{R_\ell^0}{I_\ell^0}\right)^*\tfrac{R_{\ell+1}^0}{I_{\ell+1}^0}\right] \simeq 2\omega^2  \int_{\ell_\text{cr}/\omega}^\infty db\, b \sin^2\big(\tfrac{\alpha}{2}\big),
\end{align}
where
\begin{eqnarray}
\alpha&=&\pi -2 \frac{d}{d b} \bigg[\chi(r_\text{tp})+\int_{r_\text{tp}}^\infty dr\, \Big(\tfrac{r^2+a^2}{\Delta}\Big)\Big(1-\tfrac{k}{\omega}\Big)\bigg] \nonumber \\
&=&\pi-2\int_{r_\text{tp}}^\infty \frac{dr}{\sqrt{\frac{(r^2+a^2)^2}{b^2}-\Delta}},
\end{eqnarray}
with~$r_\text{tp}$ being the larger real number satisfying 
\begin{equation}
	b=\frac{\omega(r_\text{tp}^2+a^2)}{\sqrt{r_\text{tp}^2+a^2-2M r_\text{tp}}}.
\end{equation}
Remarkably, the angle~$\alpha$ matches exactly the deflection angle of a null particle in a Schwarzschild spacetime when~$a=0$~\cite{Darwin,Misner:1974qy,BisnovatyiKogan:2008ts}; and, although we did not check it, in the case~$a\neq 0$, it is natural to expect~$\alpha$ to be the deflection angle of a null particle in a Kerr spacetime for on-axis scattering. 
It is easy to check that for large impact parameters~$b\gg M$ we recover again Einstein's deflection angle~\cite{Einstein1936LENSLIKEAO}
\begin{equation}
\alpha \simeq -\frac{4 M}{b}.
\end{equation} 
For a beam of scalar particles with maximum impact parameter~$b_{\text{max}}> 20M$ (remember that the integral in~$b$ diverges logarithmically and we need to truncate it) we find
\begin{equation}
\int_{\ell_\text{cr}/\omega}^{b_\text{max}} db\, b \sin^2\left(\frac{\alpha}{2}\right) = 4M^2\left[\Lambda^2 + \log\left(\frac{b_{\textrm{max}}}{20M}\right)\right],
\end{equation}
where the function~$\Lambda$ can be obtained numerically by performing the integration between~$\ell_\text{cr}/\omega$ and~$20 M$. This function is well fitted (accurate to 0.1\%) by
\begin{equation} \label{Lambda}
	\Lambda\simeq 1.91+0.0565\,\tilde{a}^2+0.0165\, \tilde{a}^4.
\end{equation}
Finally, putting all together (including accretion) we find that the force applied to the \acs{bh} is
\begin{equation} \label{F_quantum_rest2}
F^z \simeq -  \frac{4\pi \eta^2\hbar n}{\mu} \left[ \log\left(\frac{b_{\textrm{max}}}{20M}\right)+\frac{\ell_\text{cr}^2}{16}+\Lambda^2\right].
\end{equation} 
The quantity~$\ell_\text{cr}^2/16+\Lambda^2$ has a very mild dependence on~$\tilde{a}$, it is strictly increasing and takes values in~$[5.31,5.37]$. Here we have~$\eta^2=(2\omega M)^2 \gg1$, which is clearly in the particle limit; this is to be expected, since high-frequency modes are known to be well described by \emph{geometrical optics} (i.e., geodesics).

\section{Black hole moving through a scalar field}\label{sec:BHmoving}

Now we know the rate at which energy and linear momentum is imparted to a Kerr \acs{bh} by a scalar field scattering it off, from the point of view of a distant observer stationary with respect to the~\acs{bh} ("{\acs{bh} frame"). We would like to use this information to find out what are these rates, now from the point of view of a distant observer stationary with respect to the asymptotic scalar field ("scalar field frame"). The latter observer perceives the \acs{bh} moving with constant velocity~$\bm{v}=-\bm{\xi}k_\infty/\omega$ with respect to the asymptotic scalar field, which is perceived at rest (by definition). While at rest (and neglecting quantum effects) the~\acs{bh} is a perfect absorber, but when moving it may transfer some of its kinetic energy to the scalar field, with the interesting possibility of, globally, losing energy. The deposition of the \acs{bh}'s kinetic energy on the scalar field environment is intrinsically connected with the phenomenon of \acs{df}. Knowing the rate at which energy is accreted and linear momentum is imparted to the moving \acs{bh} allows us to compute how its relative motion with respect to the scalar field will evolve in time due to \acs{df}\footnote{In nonrelativistic treatments, \acs{df} (as standing for the gravitational interaction of a perturber with its wake) and accretion of momentum are distinct effects, both contributing to the dynamics of gravitational systems; in relativistic treatments the separation into these two effects is not absolute, but gauge-dependent instead~\cite{Clough:2021qlv,Traykova:2021dua} (see, e.g., Eq.~(36) and subsequent discussion in Ref.~\cite{Clough:2021qlv}). In this work we (abusively) call \acs{df} to the total effect, which includes also the accretion of momentum.}.

The scalar field frame is related to the \acs{bh} frame through a Lorentz boost with velocity~$-\bm{v}$. Noting that the curvature \acs{adm} four-momentum~$P_\text{BH}^\alpha$ transforms as Lorentz four-vector~\cite{Arnowitt:1962hi}, it is trivial to find the rates in the scalar field frame (primed quantities)
\begin{equation}
	\dot{E}'_\text{BH}=\dot{E}_\text{BH}+\bm{v}\cdot \bm{F},
\end{equation}
\begin{equation}\label{boost}
	\bm{F}'=\bm{F}+ \dot{E}_\text{BH}\bm{v},
\end{equation}
where we used~$dt'=dt/\sqrt{1-v^2}$ and~$\bm{F} \parallel \bm{v}$. In the general case, the force does not need to point along~$\bm{v}$ (\eg, Magnus effect), but as seen in the previous section the force does oppose the velocity in the low-frequency limit or for sufficiently small incident angles -- the cases we focused on in this work. The wave effects are controlled by the parameter~\eqref{particleness}, which, after substituting~$\omega=\mu/\sqrt{1-v^2}$ and~$k_\infty=\mu v/\sqrt{1-v^2}$, reads
\begin{equation}\label{eta}
	\eta=-\frac{\mu M(1+v^2)}{v\sqrt{1-v^2}}.
\end{equation}
%


\vskip 1cm
\subsection{Weak field regime}}

For \acs{bh} velocities satisfying~$1-v^2\gg \mu^2 M^2$ the scalar field is perceived with low frequency ($\omega M \ll1$) in the \acs{bh} frame and, so, only probes the weak (Newtonian) gravitational field, as shown in Sec.~\ref{sec:BHrest}.
This limit is possible only for light fields~$\mu M \ll1$. On the other hand, for light fields all relevant astrophysical velocities are in this regime; in other words, light fields are not expected to probe the strong gravitational field of \acs{bh}s, because their de Broglie wavelength is too large.

In the scalar field frame the rate of change of the \acs{bh}'s energy is
\begin{eqnarray}
	&&\frac{\dot{E}_\text{BH}'}{\rho}=\frac{A_\text{h}}{1-v^2} \frac{e^{-\pi \eta}\pi \eta}{\sinh(\pi \eta)} \nn \\
	&&-\frac{4 \pi \eta^2v}{\mu^2} \Re\Big[\Psi(1+\ell_\text{max}+i\eta)-\Psi(1+i\eta)\Big],
\end{eqnarray}
and the \acs{df} is
\begin{equation} \label{DF_weak}
\bm{F}'=-\frac{4 \pi \eta^2 \rho\bm{v}}{\mu^2 v}\Re\Big[\Psi(1+\ell_\text{max}+i\eta)-\Psi(1+i\eta)\Big],
\end{equation}
where we have defined the scalar's proper mass density~$\rho \equiv n m_S$ and the medium size is
\begin{equation}
	b_\text{max} =\frac{\sqrt{\ell_\text{max}(\ell_\text{max}+1)(1-v^2)}}{\mu v}.
\end{equation}
We remark that, as discussed in the previous section, the analytic approximations that we are using here can only be trusted (with an error~$< 5\%$) for~$\mu M/\sqrt{1-v^2}\leq 0.01$. 

In the particle limit~$\eta^2\gg1$, which corresponds to nonrelativistic velocities~$v\ll \mu M$, the above expressions reduce to
\begin{equation}
	\dot{E}'_\text{BH}=\frac{4 \pi M^2 \rho }{v} \left\{\frac{\mu A_\text{h}}{8 M}-\log\left(\sqrt{1+\frac{b_\text{max}^2}{(M/v^2)^2}}\right)\right\},
\end{equation}
and
\begin{equation}
	\bm{F}'=- \frac{4 \pi M^2 \rho  \bm{v}}{v^3} \log\left(\sqrt{1+\frac{b_\text{max}^2}{(M/v^2)^2}}\right).
\end{equation}
For an extended medium~$b_\text{max}\gg M/v^2$ we recover Chandrasekhar's result for \acs{df} in collisionless media~\cite{Chandrasekhar:1943v1}.

In the wave limit~$\eta^2\ll1$, which corresponds to~\acs{bh} velocities~$v\gg \mu M$, the rate of change of the \acs{bh}'s energy becomes
\begin{eqnarray}
	\dot{E}_\text{BH}'&=&\frac{4\pi M^2 \rho }{v(1-v^2)}\nn\\
	&&\times\left\{\frac{A_\text{h}v}{4\pi M^2}-(1+v^2)^2\left[\log\left(\frac{v \mu b_\text{max}}{\sqrt{1-v^2}}\right)+\gamma_\text{E}\right]\right\}, \nn \\
\end{eqnarray}
and the \acs{df}
\begin{equation}\label{DF_weak_particle}
	\bm{F}'=-\frac{4\pi M^2 \rho (1+v^2)^2 \bm{v}}{v^3(1-v^2)}\left[\log\left(\frac{v \mu b_\text{max}}{\sqrt{1-v^2}}\right)+\gamma_\text{E}\right]. 
\end{equation}
We see that the force on the~\acs{bh} is indeed a friction (\ie, it acts to decrease the absolute value of its velocity) in the entire range of~$v$. For nonrelativistic velocities this force reduces to
\begin{equation}
	\bm{F}'=-\frac{4\pi M^2 \rho \bm{v}}{v^3}\left[\log\left(v \mu b_\text{max}\right)+\gamma_\text{E}\right], 
\end{equation}
which for an extended medium~$ b_\text{max}\gg1/v\mu$ coincides with the result derived in~\cite{Hui:2016ltb,Lancaster:2019mde} and extracted numerically in~\cite{Traykova:2021dua} (up to an additive constant, which is due to a different cutoff scheme, to be discussed in Sec.~\ref{sec:discussion})\footnote{Upon the identification~$r\equiv b_\text{max}/2$ between the cutoff radius~$r$ employed in~\cite{Hui:2016ltb,Traykova:2021dua} and our impact parameter~$b_\text{max}$.}. 

\begin{figure}[h]
	\centering
	\begin{tabular}{c}
		\includegraphics[width=0.95 \linewidth]{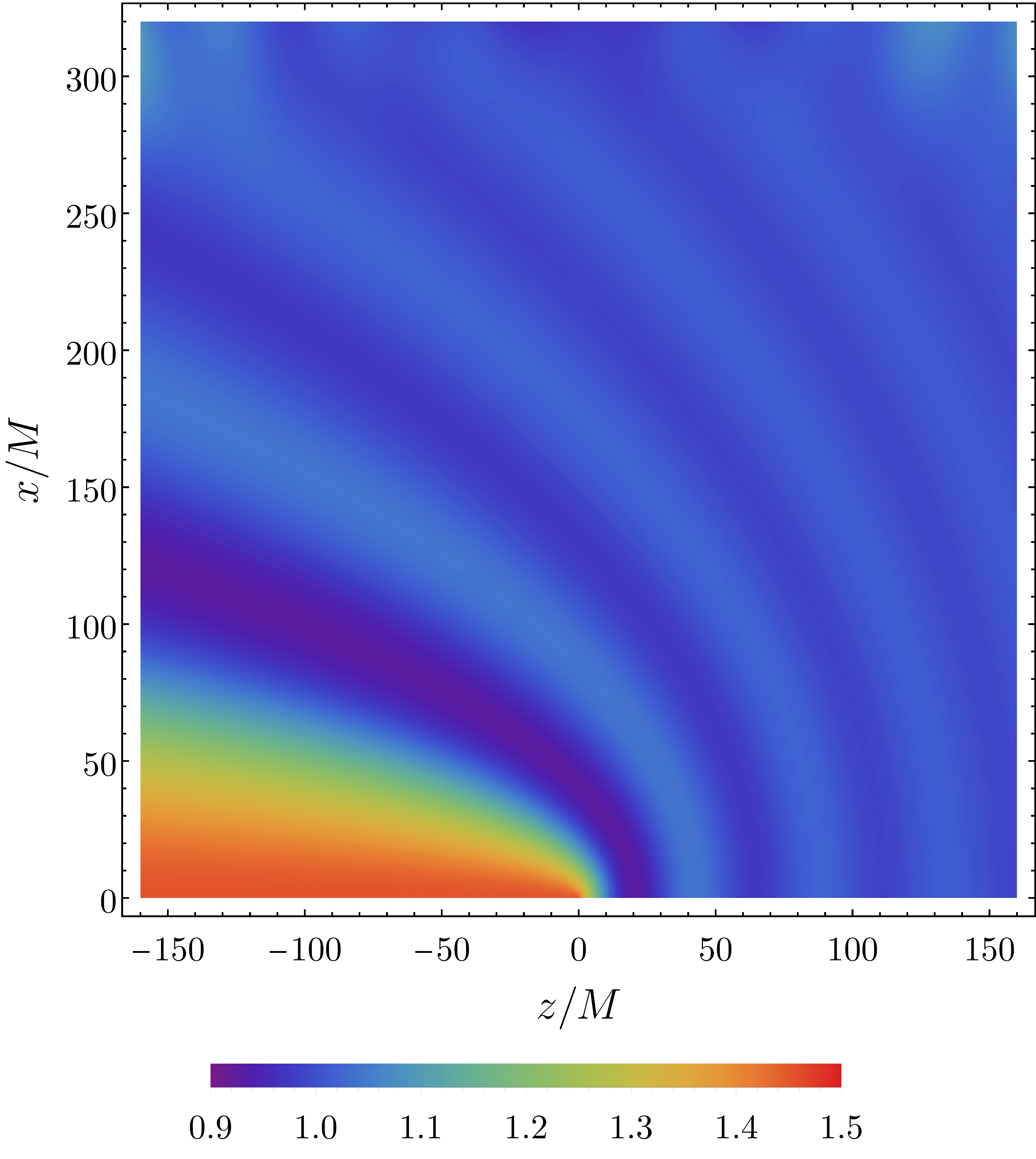} 
	\end{tabular}
	\caption{Scalar field energy density~$T_{tt}/\rho$ (in the \acs{bh} frame) in the weak field regime, obtained using the far-region solution~\eqref{scalar_weak} with coefficients~\eqref{c3} and~\eqref{c4} (in which we substitute~\eqref{T_low} and~\eqref{incident_ampl}). The~\acs{bh} is at the origin moving from left to right with velocity~$v=0.8 c$. The scalars have mass~$\mu M=0.05$ and~$b_\text{max}/M\simeq 396$. For these parameters the scalar field is close to the wave limit ($\eta^2\simeq 0.016$) as can be clearly seen by the interference fringes with characteristic length~$\sim \lambda_\text{dB} /2M$.}
	\label{fig:wake}
\end{figure}

In Fig.~\ref{fig:wake} we show the energy density~$T_{tt}/\rho$ of the scalar field in the weak field regime, obtained using the far-region solution~\eqref{scalar_weak} with coefficients~\eqref{c3} and~\eqref{c4} (in which we substitute~\eqref{T_low} and~\eqref{incident_ampl}). This wake is in the wave limit ($\eta^2\ll1$) and its wave structure (interference fringes) of  characteristic length~$\sim \lambda_\text{dB} /2M = \pi(1+v^{-2})/|\eta|$ is clearly seen. Our figure should be compared with the steady wake attained dynamically in the numerical evolution performed in Ref.~\cite{Traykova:2021dua} (showed in the second row of their Fig.~2 for the same set of parameters); the resemblance between the two images is remarkable. Going away from the wave limit into the particle limit, the fringes disappear and the wake becomes concentrated in a single tail with much greater energy density (as it was seen in~\cite{Traykova:2021dua}).


\vskip 1cm
\subsection{Strong field regime}

For velocities~$1-v^2 \ll  \mu^2 M^2$ the scalar field is perceived with high-frequency ($\omega M \gg 1$) in the \acs{bh} frame and, so, it is able to probe the strong gravity region of the \acs{bh}. Here we restrict to the special case in which the \acs{bh} motion is along the direction of its spin and we consider light scalars with mass~$\mu M\leq 1$. In this case, the condition~$1-v^2 \ll  \mu^2 M^2$ is satisfied only at ultrarelativistic speeds~$v\sim1$ and the scalar field behaves as particles ($\eta^2\gg1$).

In the scalar field frame the rate of change of the~\acs{bh}'s energy is
\begin{equation}
	\dot{E}_\text{BH}'=-\frac{16\pi M^2 \rho }{1-v^2} \left[ \log\left(\frac{b_{\textrm{max}}}{20M}\right)+\Lambda^2\right],
\end{equation}
and the \acs{df} is
\begin{equation}
	\bm{F}'=-\frac{16\pi M^2 \rho  \bm{v}}{(1-v^2)v} \left[ \log\left(\frac{b_{\textrm{max}}}{20M}\right)+\Lambda^2\right],
\end{equation}
where we recall that~$\Lambda\simeq 1.91+0.0565\tilde{a}^2+0.0165 \tilde{a}^4$ and that these expressions are valid only for~$b_\text{max}<20M$; for smaller~$b_\text{max}$ we need to perform a numerical integration.

\section{Discussion} \label{sec:discussion}

In this work we derived simple closed-form expressions for the dynamical fiction acting on \acs{bh}s moving through an ultralight scalar field, covering both nonrelativistic and relativistic speeds, and including the effect of \acs{bh} spin. We showed that for velocities~$1-v^2\gg \mu^2 M^2$ the scalar has too large a de Broglie wavelength to probe the strong gravity region of the spacetime (we called it weak field regime). In this case, for low nonrelativistic velocities~$v\ll \mu M$ the scalars behave as particles and the force on the \acs{bh} (in the scalar's frame) is
\begin{equation*}
\bm{F}'=- \frac{4 \pi M^2 \rho \bm{v}}{v^3} \log\left(\sqrt{1+\frac{b_\text{max}^2}{(M/v^2)^2}}\right).
\end{equation*}
Still in the weak field regime, the wave effects grow with the \acs{bh} velocity and are at their greatest for~$v\gg \mu M$, in which case
\begin{equation*}
\bm{F}'=-\frac{4\pi M^2 \rho (1+v^2)^2 \bm{v}}{v^3(1-v^2)}\left[\log\left(\frac{v \mu b_\text{max}}{\sqrt{1-v^2}}\right)+\gamma_\text{E}\right].
\end{equation*}
For light scalar masses~$\mu M \leq 1$ and astrophysical~\acs{bh} velocities all systems are expected to be in the weak field regime (even for the possibly relativistic velocities found in~\acs{bh} mergers). We verified that these analytic expressions describe very well the numerical results (obtained using the numerical values for the scattering amplitudes) for~$\mu M/\sqrt{1-v^2}\leq 0.01$.
For the sake of curiosity, for ultrarelativistic \acs{bh} velocities ($v\sim 1$) satisfying~$1-v^2 \ll  \mu^2 M^2$ the scalars are able to probe the strong field region of spacetime (since their de Broglie wavelength becomes much smaller than the event horizon radius) and behave again as particles; here the \acs{df} is
\begin{equation*}
\bm{F}'=-\frac{16\pi M^2 \rho  \bm{v}}{(1-v^2)v} \left[ \log\left(\frac{b_{\textrm{max}}}{20M}\right)+\Lambda^2\right].
\end{equation*}
Additionally, we derived simple expressions for the rate of change of the \acs{bh}'s energy, which allows us to do an energy balance between the kinetic energy deposited in the environment and the accreted mass; we also extended these expressions to the case of massless (scalar) radiation, covering the numerical results of Ref.~\cite{Cardoso:2019dte}.

Due to the~$1/r$ falloff of the gravitational potential, the \acs{df} in an unbounded homogeneous scalar field medium diverges and a cutoff is needed (in practice this is not a problem, because these scalar environments have a finite size, \eg, \acs{dm} halos). In previous studies (\eg, \cite{Hui:2016ltb,Lancaster:2019mde,Clough:2021qlv}) an \emph{ad hoc} cutoff scheme was employed, consisting in neglecting the contribution to \acs{df} of scalar field from a region outside a ball of radius~$r$ centered at the \acs{bh}. This is clearly not self-consistent, since the wake is computed for a medium of infinite extension. In this work we use a cutoff scheme more similar to the one employed in the original Chandrasekhar's treatment~\cite{Chandrasekhar:1943v1}, which consists in considering a maximal impact parameter for the \emph{unperturbed} medium. This approach is self-consistent since the wake is computed for the truncated medium. But, actually, there is also a subtlety with our cutoff scheme. In the \acs{bh} frame, our truncated medium is in a superposition of eigenstates of the operator~$\hat{L}^z$ with maximum eigenvalue~$\hbar \ell_\text{max}$ and with coefficients such that the expectation value of the asymptotic scalar's momentum satisfies~$\lim_{\ell_\text{max}\to \infty}\left< \hat{\bm{p}}_\infty\right >=-m_S \bm{v}/\sqrt{1-v^2}$. So, we note that the interpretation of a~\acs{bh} moving with velocity~$\bm{v}$ with respect to the scalars is only correct for~$l_\text{max}\gg 1$; in particular, in our description there is an inherent velocity dispersion~$|\Delta\bm{v}|/v\gtrsim (1-v^2)/\sqrt{\ell_\text{max}(\ell_\text{max}+1)}$, by the uncertainty principle. Interestingly, a velocity dispersion of the scalars in a \acs{dm} halo is actually expected and can be modeled through a random phase distribution, \eg, \cite{Lancaster:2019mde,Hui:2021tkt} (in principle, our framework can also be applied to such setup, but we postpone the study of this issue to future work).

For nonrelativistic \acs{bh} velocities in an extended medium of size~$b_\text{max}\gg \max\{1/v \mu,M/v^2\}$ we recover the Newtonian expressions derived in~\cite{Hui:2016ltb,Lancaster:2019mde} (up to an additive constant, which comes from the different cutoff scheme used there) and we find that the ratio of the wave to the particle \acs{df} expressions is
\begin{equation}
	\frac{F'_\text{wave}}{F'_\text{particle}}=\frac{\log\left(k_\infty b_\text{max}\right)+\gamma_E}{\log\left(k_\infty b_\text{max}\right)-\log|\eta|},
\end{equation}
which is smaller than unity in the wave limit~$|\eta|\ll1$. Remarkably, the above ratio is unchanged for relativistic velocities if for~$F_\text{particle}'$ we use the expression derived in Ref.~\cite{Petrich:1989} describing the relativistic \acs{df} in a \emph{collisionless} medium. So, we find that the wave effects of light scalars suppress \acs{df} in an extended medium, both for nonrelativistic and relativistic velocities. This fact, as remarked previously in Ref.~\cite{Hui:2016ltb}, can alleviate substantially the timing problem of the five globular clusters in the Fornax dwarf spheroidal~\cite{Tremaine:1976}, and similar issues in faint dwarfs in several nearby galaxy clusters\footnote{However, the dominant effect suppressing \acs{df} seems to be the cored density profile~\cite{Bar:2021jff,Read:2006} of the Fornax~(e.g., Fig.~6 of~\cite{Hayashi:2020jze}). These cores can arise naturally in alternative models to the standard cold \acs{dm} (like fuzzy \acs{dm}~\cite{Annulli:2020lyc,Hui:2021tkt}, but not only~\cite{Bar:2021jff}), or can develop due to baryonic feedback in cold \acs{dm} haloes~\cite{Tollet:2016}.}~\cite{Lotz:2001gz,Cowsik:2009uk}.

\begin{figure}
	\centering
	\begin{tabular}{c}
		\includegraphics[width=0.9 \linewidth]{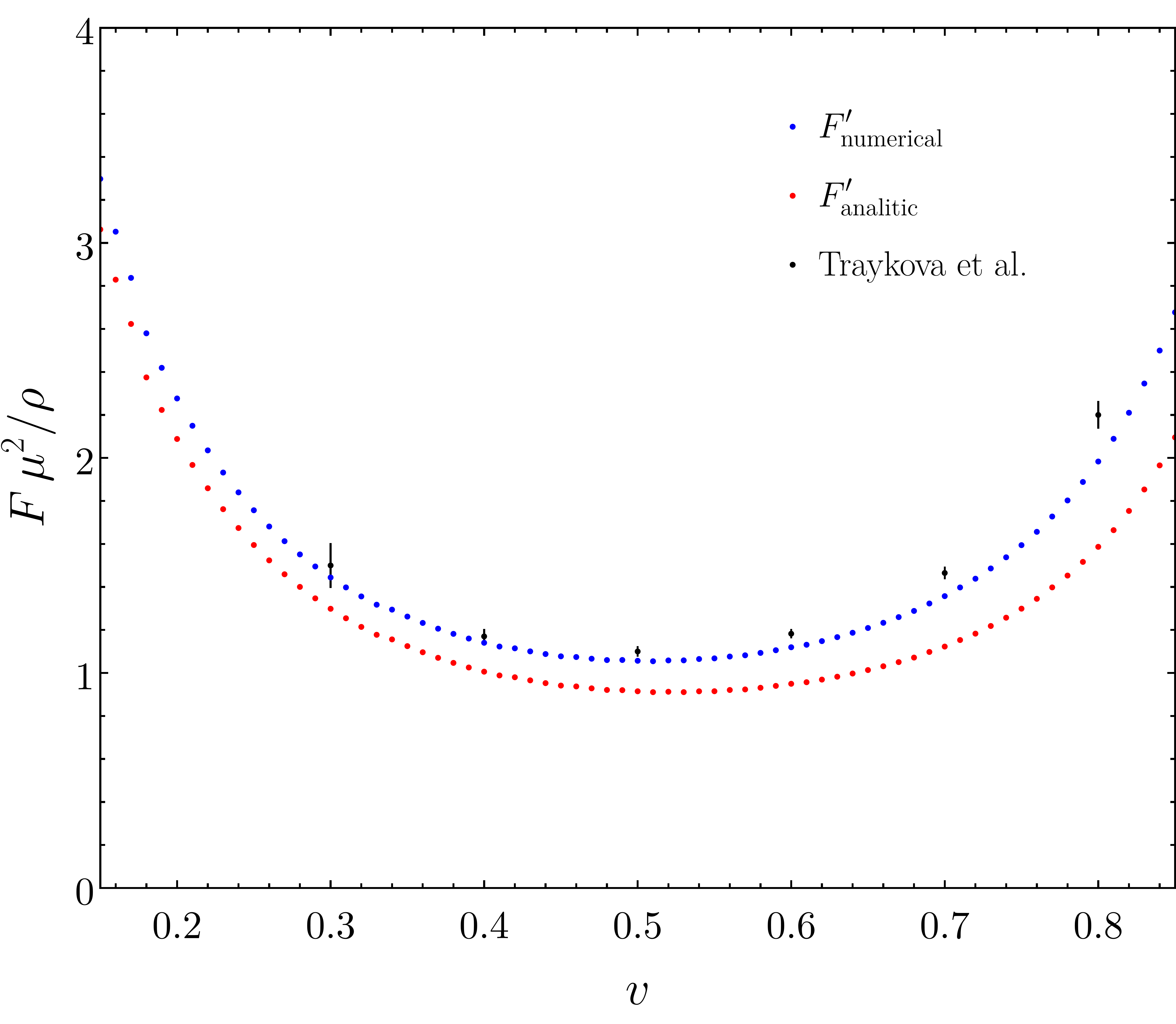} \\
		\includegraphics[width=0.9 \linewidth]{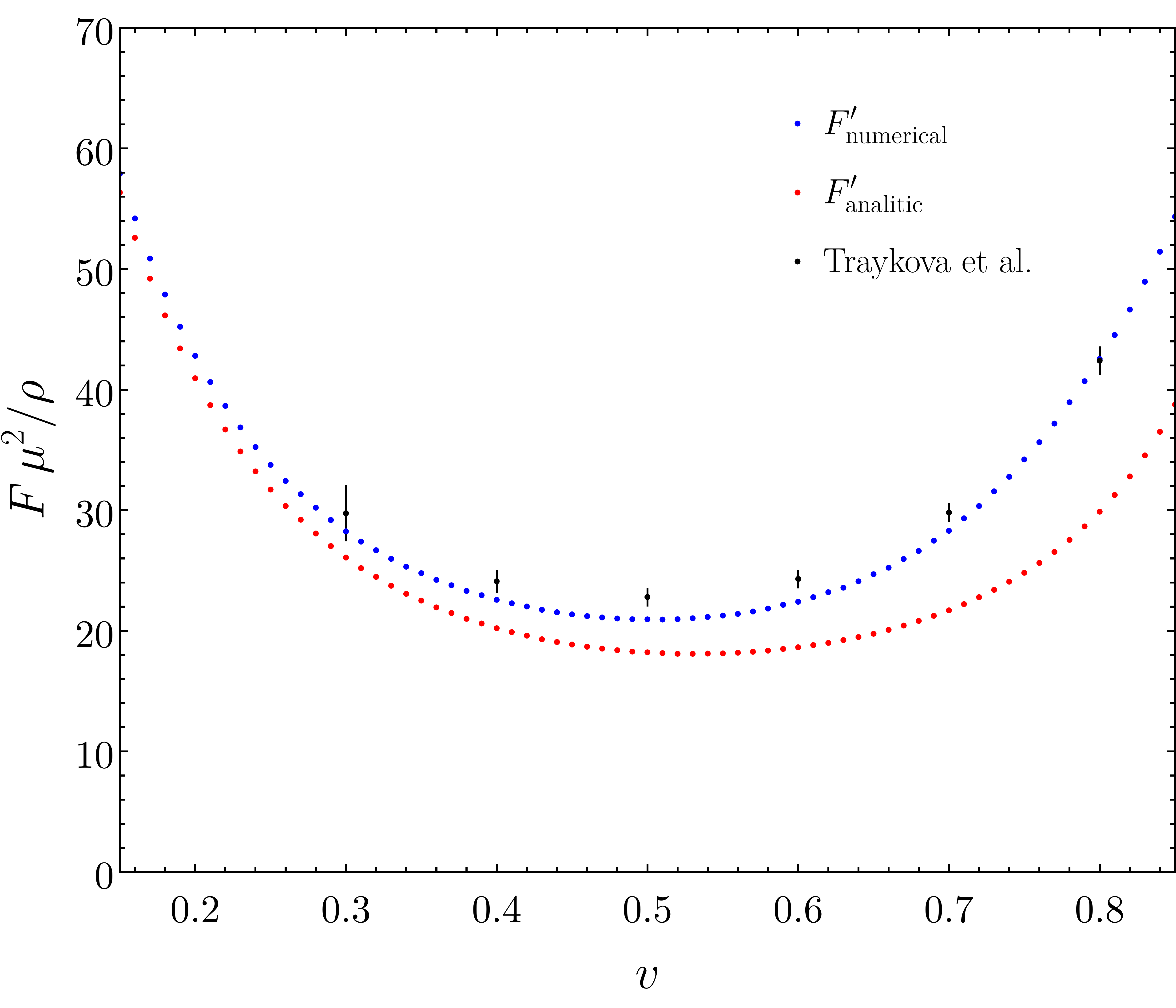}
	\end{tabular}
	\caption{Comparison between the numerical results of Traykova et al.~\cite{Traykova:2021dua} and our framework based on the scattering amplitudes with the~$R/I$ obtained numerically ($F'_\text{numerical}$) and with the analytic approximation of Eq.~\eqref{DF_weak} ($F'_\text{analitic}$); \acs{top panel}:~$\mu M=0.05$, \acs{bottom panel}:~$\mu M=0.2$. Here, the analytical approximation~\eqref{DF_weak} does not describe very well~$F'_\text{numerical}$ because~$\mu M/\sqrt{1-v^2}$ is not sufficiently small.}
	\label{fig:trayk}
\end{figure}

As inferred in the numerical treatment of Ref.~\cite{Traykova:2021dua}, we find that the relativistic corrections to \acs{df} introduce a factor~$(1+v^2)^2/(1-v^2)$; the same correction was found in~\cite{Petrich:1989} for collisionless and in~\cite{Barausse:2007ph} for collisional media. But we argue here that (at least in the weak field regime~$1-v^2\gg \mu^2 M^2$) this sole factor encodes the \emph{entire} correction to \acs{df} and that the extra corrections introduced in~\cite{Traykova:2021dua} are slightly misguided. When boosting from their \emph{simulation frame} to the \acs{bh}'s, the authors neglected the contribution of accretion (their~Eq.~$10$); actually, this contribution is important and it cannot be neglected. Because of that, their results for the \acs{df} are valid in their simulation coordinates -- and not in the \acs{bh}'s frame. It is easy to show that the force in their simulation frame is equal to the force~$\bm{F}'$ in the scalar's frame. 
As we have shown the accretion of momentum cancels out of~$\bm{F}'$ (c.f.,~\eqref{boost}). So, we argue here that both their "pressure correction" (depending on a parameter~$\kappa$) and Bondi's momentum accretion are actually describing strong gravity corrections to the \acs{df}; remember that the weak field analytic expressions that we derived are a good approximation only for~$\mu M/\sqrt{1-v^2}\leq 0.01$, for larger~$\mu M$ strong gravity effects start to kick in. We predict that these corrections will not be needed to fit their results for~$\mu M/\sqrt{1-v^2}\leq 0.01$. 
Our suspicions are supported by the fact that using our framework to compute the force~$\bm{F}'$ with the scattering amplitudes~$R/I$ obtained numerically, gives results in remarkable agreement with Ref.~\cite{Traykova:2021dua}\footnote{To do the comparison we introduced an additive constant that accounts for the different cutoff schemes (as explained before).} (as can be seen in Fig.~\ref{fig:trayk}). The main difference between that numerical procedure and the analytic expression~\eqref{DF_weak} are strong gravity corrections (which are suppressed for~$\mu M/\sqrt{1-v^2}\leq 0.01$).
Summarizing, we conclude that in the scalar field frame and in the weak field regime the relativistic corrections to \acs{df} are encoded solely in the factor~$(1+v^2)^2/(1-v^2)$.

For simplicity, in this work we considered \emph{complex} scalars, which can arise in simple extensions of the Standard Model~\cite{Freitas:2021cfi}, but this framework can also be applied to (the more physically motivated) real scalars. In that case, looking at the form of the energy-momentum tensor~\eqref{scalarEMT}, it is easy to conclude that the scalar's energy (momentum) density cannot reach a stationary state, but instead it will be left oscillating with frequency~$2 \omega M=2\mu M/\sqrt{1-v^2}$. The \acs{df} will also oscillate with the same frequency and it is straightforward to show that its average is half of the value of \acs{df} in the complex case. The same conclusion was also obtained using numerical simulations in~\cite{Traykova:2021dua}. 

In this work we included also the effect of \acs{bh} spin in \acs{df}. In the weak field regime (the most relevant for astrophysical applications) the spin does not affect~\acs{df} in the scalar field frame, and affects accretion only mildly by changing the event horizon area. In the particular case of a \acs{bh} moving at ultrarelativistic speeds with its spin aligned with the direction of motion, the \acs{df} is also almost not affected by the \acs{bh} spin. The strong field regime with a \acs{bh} spin not aligned with its direction of motion was not studied here; this is the case in which we expect the \acs{bh} spin to affect the most \acs{df}. In particular, \acs{df} will not be in general aligned with the direction of motion and there will be a Magnus effect bending the \acs{bh}'s trajectory. We postpone the study of this and other interesting phenomena to future work.

\begin{acknowledgments}
\noindent
We thank Katy Clough, Pedro Ferreira and Dina Traykova for helpful discussions about their work and for helping us showing the consistency between our results.  We also thank Emanuele Berti, Diego Blas, Miguel Correia and Ricardo Z. Ferreira for their comments. 
R.~V. was supported by "la Caixa" Foundation grant no. LCF/BQ/PI20/11760032 and Agencia Estatal de Investigación del Ministerio de Ciencia e Innovación grant no. PID2020-115845GB-I00. R. V. also acknowledges support by grant no. CERN/FIS-PAR/0023/2019. 
V. C. is a Villum Investigator supported by VILLUM FONDEN (grant no. 37766) and a DNRF Chair, funded by the Danish National Research Foundation.
V. C. acknowledges financial support provided under the European Union's H2020 ERC 
Consolidator Grant ``Matter and strong-field gravity: New frontiers in Einstein's 
theory'' grant agreement no. MaGRaTh--646597.
This project has received funding from the European Union's Horizon 2020 research and innovation programme under the Marie Sklodowska-Curie grant agreement No 101007855.
We thank FCT for financial support through Project~No.~UIDB/00099/2020.
We acknowledge financial support provided by FCT/Portugal through grants PTDC/MAT-APL/30043/2017 and PTDC/FIS-AST/7002/2020.
IFAE is partially funded by the CERCA program of the Generalitat de Catalunya.
\end{acknowledgments}

\appendix

\section{Scattering amplitudes in the low-frequency limit (\boldmath$\omega M \ll 1$)}\label{app:LF}

Here we use the method of matched asymptotic expansions to find an approximate analytic expressions for the amplitudes~$R/I$ and~$T/I$ of massive scalar waves scattering off spinning \acs{bh}s in the low-frequency limit~$\omega M \ll 1$. This is an extension of the treatment in Refs.~\cite{Starobinski:1973,Starobinski2:1973} and~\cite{Unruh:1976fm}. 

Note that the spheroidal eigenvalues have a power-expansion~\cite{NIST:DLMF}
\begin{equation}
	\lambda_\ell^m = \ell(\ell+1)+\frac{(k_\infty a)^2}{2}\bigg[1+\frac{(2m-1)(2m+1)}{(2\ell-1)(2\ell+3)}\bigg]+ \mathcal{O}[(k_0a)^4]
\end{equation}
and, so, in the low-frequency limit~$\omega M \ll 1$, the eigenvalues are~$\lambda_\ell^m\simeq \ell(\ell+1)$ at leading order.

\vskip 1cm
\subsection{Region I}
Let us consider first the region
\begin{equation}
	x\equiv\frac{r-r_\text{h}}{r_\text{h}-r_\text{c}}\ll \frac{\ell+1}{\omega (r_\text{h}-r_\text{c})},
\end{equation}
where~$r_\text{c}=M-\sqrt{M^2-a^2}$ is the radius of the Cauchy horizon (the smallest real root of~$\Delta$). In this region Eq.~\eqref{KG_rad_alt} reduces to~\cite{Starobinski:1973}
\begin{equation}
	x(x+1)\frac{d}{dx}\left[x(x+1)\frac{d \mathcal{R}}{dx}\right]+\big[Q^2-\ell(\ell+1)x(x+1)\big]\mathcal{R}=0,
\end{equation}
where
\begin{equation}
	Q=\frac{r_\text{h}^2+a^2}{r_\text{h}-r_\text{c}}(m \Omega_\text{h}-\omega).
\end{equation}
The general solution of this equation is~\cite{NIST:DLMF}
\begin{eqnarray}
	\mathcal{R}&=&\left(1+x\right)^{iQ}\bigg\{c_1 x^{-iQ}\,\text{F}(-\ell,\ell+1;1-\bar{Q};-x) \nn \\
	&&+c_2 x^{iQ}\,\text{F}(-\ell+\bar{Q},\ell+1+\bar{Q};1+\bar{Q};-x) \bigg\},\nn\\
\end{eqnarray}
with~$\bar{Q}\equiv 2iQ$ and where~$\text{F}(a,b;c;z)$ is the hypergeometric function~\cite{NIST:DLMF}. 
The physical boundary conditions~\eqref{scalar_horiz} at the event horizon~($x\to 0^{+}$) imply that
\begin{equation}
	c_1=0,
\end{equation}
\begin{equation}
	c_2=\frac{T}{(r_\text{h}^2+a^2)^{\frac{1}{2}}}.
\end{equation}
Note that the tortoise coordinate~$\chi$ in~\eqref{scalar_horiz} is defined up to an additive constant that we have fixed here through the condition~$\chi(r)\simeq \left(\frac{r_\text{h}^2+a^2}{r_\text{h}-r_\text{c}}\right)\log\left(\frac{r-r_\text{h}}{r_\text{h}-r_\text{c}}\right)$ for~$r\sim r_\text{h}$.

In the limit~$x\gg 1$ one finds that the above solution has the form~\cite{Starobinski2:1973}
\begin{equation} \label{asympI}
	\mathcal{R}\simeq d_1 x^\ell+ \left(\frac{d_2}{2 \ell+1}\right)x^{-\ell-1},
\end{equation}
with
\begin{equation}
	d_1=T\left[\frac{(2\ell)!}{\ell!\, (1+\bar{Q})_\ell \,(r_\text{h}^2+a^2)^{\frac{1}{2}}}\right],
\end{equation}
\begin{eqnarray}
	d_{2}=(-1)^{\ell+1}T \left[\frac{\ell!\,(\bar{Q}-\ell)_{\ell+1}}{2(2 \ell)!\,(r_\text{h}^2+a^2)^{\frac{1}{2}}}\right] ,
\end{eqnarray}
where~$(z)_n\equiv z(z+1)\cdots(z+n-1)$ is the Pochhammer symbol.

\vskip 1cm
\subsection{Region II}
Now we focus on the region~$r\gg r_\text{h}$ (which implies $x\gg 1$), where Eq.~\eqref{KG_rad_alt} reduces to 
\begin{equation} \label{radialEOMII}
	\left[\frac{d^2}{d r^2}+k_\infty^2-\frac{2\eta k_\infty}{r}-\frac{\ell(\ell+1)}{r^2}\right]\left(\sqrt{\Delta}\, \mathcal{R}\right)=0.
\end{equation}
To obtain the last equation we neglected terms of order~$(r_\text{h}/r)^{3}$, used the low-frequency condition $\omega M \ll 1$, and defined the parameter
\begin{equation}
	\eta\equiv-M\left(\frac{\omega^2+k_\infty^2}{k_\infty}\right).
\end{equation}
This equation admits the solution
\begin{equation} \label{scalar_weak}
	\mathcal{R}=\frac{c_3}{r}\,\text{F}_\ell^C(\eta, k_\infty r)+\frac{c_4}{r}\,\text{G}_\ell^C(\eta, k_\infty r),
\end{equation}
where~$\text{F}_\ell^C$ and~$\text{G}_\ell^C$ are the Coulomb wave functions~\cite{NIST:DLMF}.

At~$k_\infty r \ll l$ the solution has the polynomial form~\cite{NIST:DLMF}
\begin{equation} \label{asympII}
	\mathcal{R} \simeq c_3 C_\ell(\eta)k_\infty^{\ell+1}r^\ell+c_4\frac{k_\infty^{-\ell} r^{-\ell-1}}{(2\ell+1)C_\ell(\eta)},
\end{equation}
with
\begin{equation}
	C_\ell=\frac{2^\ell e^{-\eta \pi/2} |\Gamma(\ell+1+i \eta)|}{(2\ell+1)!}.
\end{equation}

At spatial infinity~$k_\infty r \to \infty$ it has the asymptotic form
\begin{equation}
	\mathcal{R} \simeq \frac{c_3}{r} \sin [\theta_\ell(\eta, k_\infty r)]+ \frac{c_4}{r} \cos [\theta_\ell(\eta, k_\infty r)],
\end{equation}
where
\begin{equation}
	\theta_\ell=k_\infty r- \eta \log(2 k_\infty r)-\ell\frac{\pi}{2}+ \arg \Gamma(\ell+1+i \eta),
\end{equation}
with~$\arg(z)$ the principal argument of~$z$. The physical boundary conditions~\eqref{scalar_infin} imply that
\begin{equation}\label{asymp_incid}
	I=\left(\frac{c_4+i c_3}{2}\right)e^{i\left[\ell \pi/2-\arg \Gamma(\ell+1+i \eta)\right]},
\end{equation}
\begin{equation}\label{asymp_reflect}
	R=\left(\frac{c_4-i c_3}{2}\right)e^{-i\left[\ell \pi/2-\arg \Gamma(\ell+1+i \eta)\right]}.
\end{equation}
%
\vskip 1cm
\subsection{Matching the two regions}
Finally, we just need to match the solutions in the two regions. Matching~\eqref{asympII} with~\eqref{asympI} at~$r_\text{h}\ll r \ll 1/k_\infty$ gives
\begin{equation}\label{c3}
c_3=T\bigg[\frac{(2 \ell)!\,(M^2-a^2)^{-\frac{\ell}{2}}}{2^{\ell}\,\ell!\,(1+\bar{Q})_l \,(r_\text{h}^2+a^2)^{1/2}C_\ell\, k_\infty^{\ell+1}}\bigg],
\end{equation}
\begin{equation}\label{c4}
c_4=(-1)^{\ell+1}\, T\bigg[\frac{2^{\ell}\,\ell!\,(\bar{Q}-\ell)_{\ell+1}\,(M^2-a^2)^{\frac{\ell+1}{2}}}{(2 \ell)!\,(r_\text{h}^2+a^2)^{1/2}C_\ell^{-1}k_\infty^{-\ell}}\bigg] .
\end{equation}
So, using Eqs.~\eqref{asymp_incid} and~\eqref{asymp_reflect} we find the following scattering amplitudes:
\begin{eqnarray} \label{R_low}
&&\frac{R}{I}=(-1)^{\ell+1}e^{2i \arg \Gamma(\ell+1+i \eta)}\times \nn\\
&&\Bigg\{\frac{[(2\ell)!]^2+Q\,(\ell!)^2 C_\ell^2 \,|(1+2iQ)_\ell|^2\big(2 k_\infty \sqrt{M^2-a^2}\,\big)^{2 \ell+1}}{[(2\ell)!]^2-Q\,(\ell!)^2 C_\ell^2 \,|(1+2i Q)_\ell|^2\big(2 k_\infty \sqrt{M^2-a^2}\,\big)^{2 \ell+1}}\Bigg\},\nn\\ 
\end{eqnarray}
\begin{eqnarray}\label{T_low}
&&\frac{T}{I}= (-i)^{\ell+1} e^{i \arg \Gamma(\ell+1+i \eta)}\left(\frac{k_\infty }{|\omega-m \Omega_\text{h}|}\right)^\frac{1}{2} \times\nn \\
&& \Bigg\{\frac{2\sqrt{|Q|}\,\ell!\, (2l)! \,C_\ell\,	(1+2iQ)_\ell\, \big(2 k_\infty \sqrt{M^2-a^2}\,\big)^{ \ell+\frac{1}{2}}}{[(2\ell)!]^2-Q\,(\ell!)^2 C_\ell^2 \,|(1+2i Q)_\ell|^2\big(2 k_\infty \sqrt{M^2-a^2}\,\big)^{2 \ell+1}}\Bigg\}.\nn \\
\end{eqnarray}
It is easy to verify that the last expressions satisfy the conservation of the Wronskian~\eqref{wronsk}.

For a static (Schwarzschild) \acs{bh} we have~$a=0$,~$r_\text{h}=2M$ and~$Q=2\omega M \ll 1$, the last expressions simplify to
\begin{eqnarray}
\frac{R}{I}&=&(-1)^{\ell+1}e^{2i \arg \Gamma(\ell+1+i \eta)}\nn \\
&&\times\Bigg\{\frac{[(2\ell)!]^2-(\ell!)^4 C_\ell^2 \,\big(2 k_\infty M\,\big)^{2( \ell+1)}(\omega/k_\infty)}{[(2\ell)!]^2+(\ell!)^4 C_\ell^2 \,\big(2 k_\infty M\,\big)^{2( \ell+1)}(\omega/k_\infty)}\Bigg\}  \nn \\ \nn \\
&\simeq&(-1)^{\ell+1}e^{2i \arg \Gamma(\ell+1+i \eta)}\nn \\
&&\times\Bigg\{1-\left(\frac{2(\ell!)^4}{[(2\ell)!]^2}\right)\left(\frac{\omega}{k_\infty}\right) C_\ell^2 \,\big(2 k_\infty M\,\big)^{2( \ell+1)}\Bigg\},\nn \\
\end{eqnarray}
\begin{eqnarray}
\frac{T}{I}&=&(-i)^{\ell+1} e^{i \arg \Gamma(\ell+1+i \eta)} \nn \\
&&\times\Bigg\{\frac{2\,(\ell!)^2\, (2l)! \,C_\ell\, \big(2 k_\infty M\,\big)^{ \ell+1}}{[(2\ell)!]^2+(\ell!)^4 C_\ell^2 \,\big(2 k_\infty M\,\big)^{2( \ell+1)}(\omega/k_\infty)}\Bigg\} \nn \\ \nn \\
&\simeq&(-i)^{\ell+1} e^{i \arg \Gamma(\ell+1+i \eta)} \left(\frac{2\,(\ell!)^2}{(2\ell)!}\right) C_\ell\, \big(2 k_\infty M\,\big)^{ \ell+1}, \nn \\
\end{eqnarray}
which agrees with previous calculations~\cite{Unruh:1976fm,Vicente:2021mqd}. 

Note that this method does not assume~$Q\ll1$ in the case of a spinning \acs{bh}, and the expressions for the amplitudes hold for any~$Q$ (as long as~$\omega M \ll 1$). Since the derivation assumes~$\omega M\ll1$,
the expressions~\eqref{R_low} and~\eqref{T_low} can be written in the simpler form
\begin{eqnarray} \label{R_lower}
&&\frac{R}{I}=(-1)^{\ell+1}e^{2i \arg \Gamma(\ell+1+i \eta)} \times\nn \\
&&\Bigg\{1+\left(\frac{2(\ell!)^2}{[(2\ell)!]^2}\right)Q\, C_\ell^2 \,|(1+2iQ)_\ell|^2\big(2 k_\infty \sqrt{M^2-a^2}\,\big)^{2 \ell+1}\Bigg\}, \nn \\
\end{eqnarray}
\begin{eqnarray}\label{T_lower}
\frac{T}{I}&=& (-i)^{\ell+1} e^{i \arg \Gamma(\ell+1+i \eta)}\left(\frac{2\,\ell!}{(2\ell)!}\right)\left(\frac{k_\infty }{|\omega-m \Omega_\text{h}|}\right)^\frac{1}{2}\nn \\
&\times&\sqrt{|Q|} \,C_\ell\,	(1+2iQ)_\ell\, \big(2 k_\infty \sqrt{M^2-a^2}\,\big)^{ \ell+\frac{1}{2}}.
\end{eqnarray}
To see this one should note that: (i)~when~$|Q| \to \infty$,  $Q|(1+2iQ)_\ell|^2\sim \mathcal{O}\big[1/(M^2-a^2)^{\ell+\frac{1}{2}}\big]$; (ii)~when~$\eta\to-\infty$, $C_\ell^2 \sim \mathcal{O}\big[(M \mu^2/k_\infty)^{2 \ell+1}\big]$, which can be seen more easily through the alternative form of~$C_\ell$~\cite{NIST:DLMF}:
\begin{equation}
	C_\ell=\frac{2^\ell \left\{\big[2\pi \eta/\big(e^{2 \pi \eta}-1\big)\big] \prod_{j=1}^\ell \left(\eta^2+j^2\right)\right\}^{1/2}}{(2\ell+1)!}.
\end{equation}

\section{Scattering amplitudes in the high-frequency limit (\boldmath$\omega M \gg 1$)}
\label{app:HF}
In the high-frequency limit we will focus only on the ultrarelativistic regime~$\omega \gg \mu$ (which, in particular, is the only possibility for scalars with~$\mu M\leq1$). This limit in frequency was studied for instance in Refs.~\cite{Sanchez:1976,Andersson:1995vi,Glampedakis:2001cx}. For very large azimuthal numbers~$\ell \gg \omega M$ using the \acs{wkb} approximation to solve Eq.~\eqref{KG_rad}, with the physical boundary conditions~\eqref{scalar_infin} and~\eqref{scalar_horiz}, one finds that~\cite{landau1981quantum}
\begin{equation} \label{WKB}
\frac{R}{I} = i \exp \Bigg\{-2i\omega\bigg[\chi(r_\text{tp})+\int_{r_\text{tp}}^\infty dr\, \Big(\tfrac{r^2+a^2}{\Delta}\Big)\Big(1-\tfrac{k}{\omega}\Big)\bigg]\Bigg\},
\end{equation}
where
\begin{equation}
k(r) = \left[\bigg(\omega-\frac{m a}{r^2+a^2} \bigg)^2-\frac{\Delta}{(r^2+a^2)^2}\left(\ell+\tfrac{1}{2}\right)^2\right]^\frac{1}{2},
\end{equation}
and~$r_\text{tp}$ is the largest classical turning point satisfying~$k(r_\text{tp})=0$, with the tortoise coordinate fixed by the condition~$\chi(r)	\simeq r+ 2 M \log(2\omega r)$ for~$r\gg r_\text{h}$; in particular,
\begin{eqnarray}
	\chi(r_\text{tp})&=&r_\text{tp}-\left(\frac{r_\text{c}^2+a^2}{r_\text{h}-r_\text{c}}\right)\log\left[2 \omega(r_\text{tp}-r_\text{c})\right] \nn \\
	&&+\left(\frac{r_\text{h}^2+a^2}{r_\text{h}-r_\text{c}}\right)\log\left[2 \omega(r_\text{tp}-r_\text{h})\right].
\end{eqnarray}

For large azimuthal numbers~$\ell\sim \omega M$ it is also possible to use the \acs{wkb} approximation to compute the absolute value~\cite{landau1981quantum}
\begin{equation}
\left|\frac{R}{I}\right|^2\simeq \frac{1}{1+ e^{2\pi \epsilon}},
\end{equation}
where
\begin{equation}
	\epsilon=\frac{k_{\min}^2}{\sqrt{2 \left(\frac{d^2}{d \chi^2} k^2\right)_{\chi_{\min}}}}
\end{equation}
with~$k_{\min}=k(\chi_{\min})$ and where~$\chi_{\min}\left(\frac{\ell}{\omega M},\frac{m}{\ell}\right)$ is the largest (real) root of
\begin{equation}
	\left(\frac{d}{d \chi} k^2\right)_{\chi_{\min}}=0.
\end{equation}
Note that, in the large~$\ell$ limit,~$\chi_{\min}$ is indeed only a function of the ratios~$\ell/\omega M$ and~$m/\ell$. Although not easy to show explicitly for a general~$a$, in the high-frequency limit~$\omega M \gg 1$ we expect~$\epsilon$ to be a monotonic rapidly decreasing function of~$\ell/\omega M$, crossing zero at a critical~$(\ell/\omega M)_\text{cr}$ which is a function of~$m/\ell$ and~$a/M$. This expectation is motivated by what happens for~$a=0$, in which case~$(\ell/\omega M)_\text{cr}=3\sqrt{3}$ and 
\begin{equation}
\epsilon=\frac{27(\omega M)^2}{2\ell}\left[1-\left(\frac{\ell}{3 \sqrt{3}\,\omega M}\right)^2\right]
\end{equation} 
and was confirmed by our numerics.
Thus, one concludes that in the high-frequency limit the reflectivity~$|R/I|^2$ is well-approximated by a very steep function of~$\ell/\omega M$ that vanishes for~$\ell/\omega M<(\ell/\omega M)_\text{cr}$ and is unity for $(\ell/\omega M)>(\ell/\omega M)_\text{cr}$. One is, then, led to the (geometrical optics) approximation~\cite{FORD1959259}
\begin{equation} \label{ratio_high_freq}
\frac{R}{I}=\begin{cases}
&0 ,\quad  \ell< \ell_\text{cr} \\
&i e^ {-2i\omega\big[\chi(r_\text{tp})+\int_{r_\text{tp}}^\infty dr\, \big(\tfrac{r^2+a^2}{\Delta}\big)\big(1-\tfrac{k}{\omega}\big)\big]}, \quad \ell\geq \ell_\text{cr}
\end{cases}. 
\end{equation}

\begin{figure}
	\centering
	\begin{tabular}{c}
		\includegraphics[width=0.95 \linewidth]{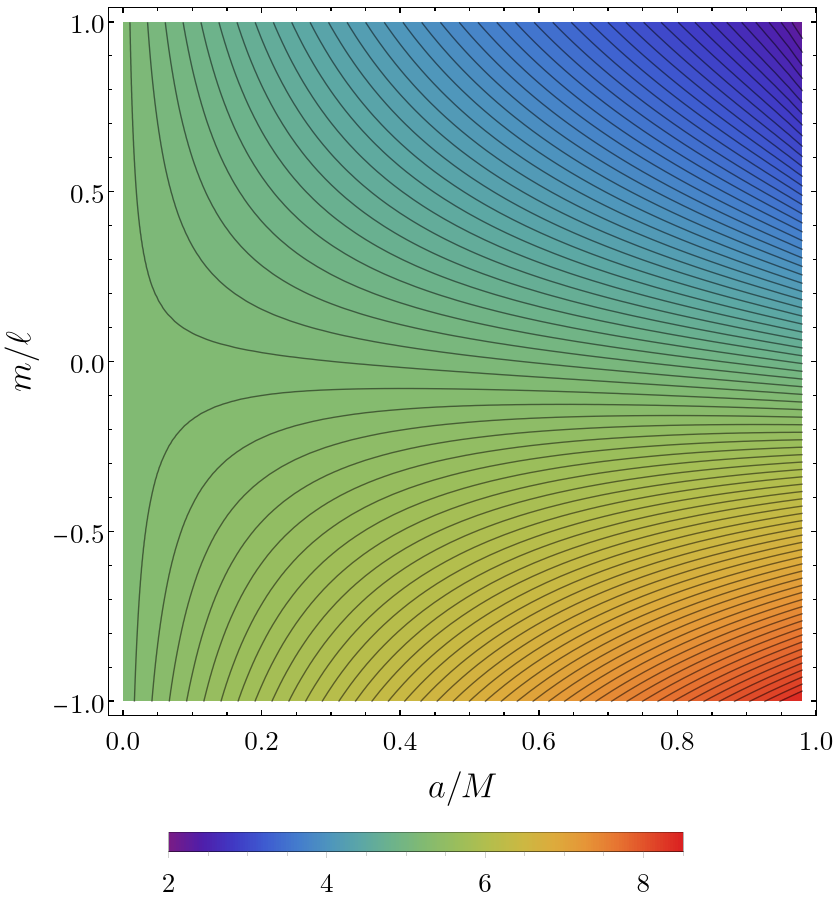} 
	\end{tabular}
	\caption{Numerical result of~$(\ell/\omega M)_\text{cr}$ as function of~$a/M$ and~$m/\ell$.}
	\label{fig:lcr}
\end{figure}

\begin{figure}
	\centering
	\begin{tabular}{c}
		\includegraphics[width=0.95 \linewidth]{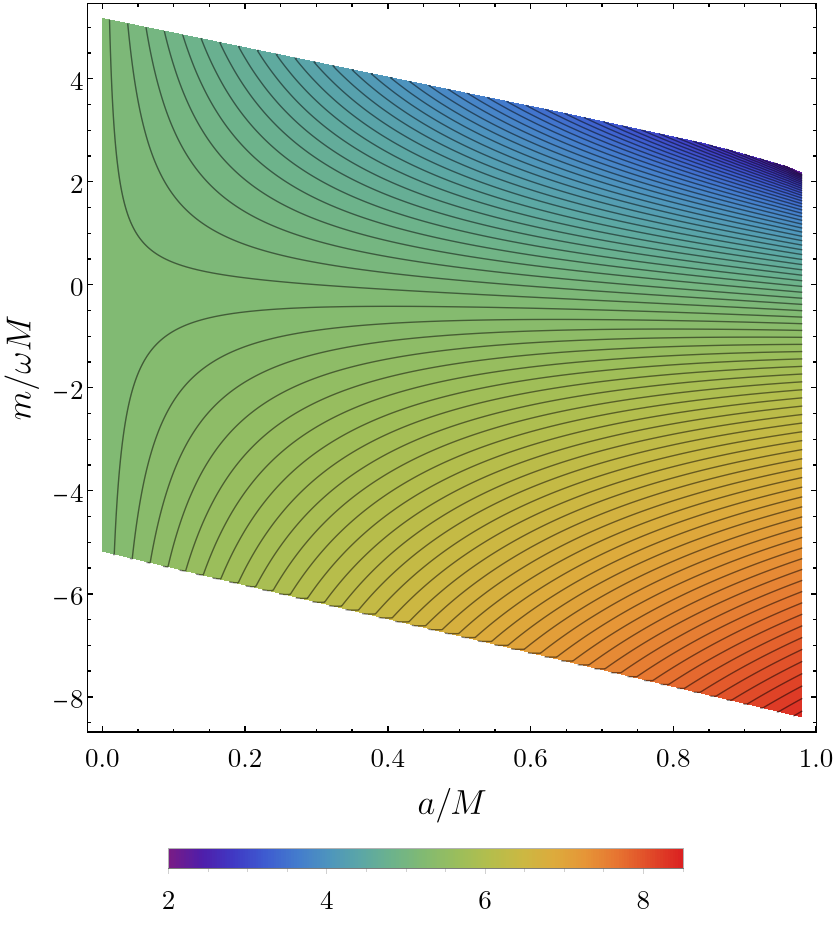} 
	\end{tabular}
	\caption{Numerical result of~$(\ell/\omega M)_\text{cr}$ as function of~$a/M$ and~$m/\omega M$. In the (unbounded) white region of parameter space there exists no~$(\ell/\omega M)_\text{cr}$ and expression~\eqref{WKB} can be used for any~$\ell/\omega M$ (as long as~$|m|\leq l$).}
	\label{fig:lcr2}
\end{figure}

Note that~$(\ell/\omega M)_\text{cr}$ is a root of the discriminant of~$k^2(r)$. This discriminant is a polynomial of degree~$8$ in~$\omega M/l$ with a double root at zero, two complex roots, two negative and two positive roots (this was established by our numerics for the physical parameters~$0\leq a/M \leq 1$ and~$-1\leq m/l \leq 1$). We have shown numerically that~$(\ell/\omega M)_\text{cr}$ is always the largest real~$\ell/\omega M$ that is a root of~$k^2(r)$. This gives us a very efficient way to compute numerically~$(\ell/\omega M)_\text{cr}$ as function of~$m/\ell$ and~$a/M$ (shown in~Fig.~\ref{fig:lcr}). Alternatively, one can use an analogous procedure to compute numerically~$(\ell/\omega M)_\text{cr}$ as function of~$m/\omega M$ and~$a/M$ (shown in~Fig.~\ref{fig:lcr2}).
%

\section{Black hole moving through a massless scalar field}\label{sec:BHmoving_massless}

The problem of obtaining the rate of change of the \acs{bh}'s energy as it moves through a massless radiation (scalar) field was solved numerically in Ref.~\cite{Cardoso:2019dte}. There, it was concluded that, depending on the medium size, there is a critical velocity above which the \acs{bh} deposits kinetic energy in its environment at a greater rate than it accretes. In this appendix we show that, when moving through a massless scalar field, both the rate of change of the \acs{bh}'s energy and the \acs{df} acting on it can be computed analytically.

In the case of a massless scalar field we cannot go to its proper frame. So, we consider here a "lab frame" with respect to which the \acs{bh} moves at velocity~$\bm{v}$ and the massless scalar has momentum~$-\hbar \omega' \bm{v}/v$. So, we have~$\omega=\sqrt{\frac{1+v}{1-v}}\omega'$ and~$\bm{k}=-\sqrt{\frac{1+v}{1-v}}\omega' \bm{v}/v$ in the~\acs{bh} frame\footnote{We could also consider a more general setup in which the \acs{bh} is not moving \emph{head-on} against the scalar field. But this particular setup is specially interesting, because when the~\acs{bh} moves at ultrarelativistic speeds ($v\sim1$), due to relativistic beaming, even isotropic radiation is perceived as counter-moving in the \acs{bh} frame~\cite{Cardoso:2019dte}. However, we do not expect to find \acs{bh}s moving at ultrarelativistic speeds in our Universe.}. The factor~$n \omega'/\mu$ is not well defined here (remember that~$n$ is the number density far from the \acs{bh} in the scalar's proper frame); this factor must be replaced by the number density in the lab frame~$n'$ (as can be readily seen by continuity, taking the limit~$m_S\to0$).

\vskip 1cm
\subsection{Weak field regime}

For \acs{bh} velocities satisfying~$1-v \gg \omega'^2M^2$ the scalar field only probes the weak (Newtonian) gravitational field. In this regime the rate of change of the \acs{bh}'s energy is
\begin{eqnarray}
&&\dot{E}'_\text{BH}=\frac{16 \pi M^2n'\hbar \omega'(1+v)v}{1-v}\nn \\
&&\times \left\{\frac{A_\text{h}}{16 \pi M^2}\left(\frac{1}{v}-1\right)- \log\left(\sqrt{\frac{1+v}{1-v}}\omega' b_\text{max}\right)- \gamma_ \text{EM}\right\},\nn \\
\end{eqnarray}
and the \acs{df} is
\begin{eqnarray}
	(\bm{F})_\text{lab}&\simeq&-\frac{16 \pi M^2 n'\hbar \omega'(1+v)\bm{v}}{(1-v)v}\nn \\
	&&\times \left\{ \log\left(\sqrt{\frac{1+v}{1-v}}\omega' b_\text{max}\right)+ \gamma_ \text{EM}+\frac{A_\text{h}(1-v)}{16 \pi M^2}\right\}.\nn \\
\end{eqnarray}

\vskip 1cm
\subsection{Strong field regime}
%
For velocities~$1-v \ll \omega'^2M^2$ the scalar field is perceived with high frequency ($\omega M \gg 1$) in the \acs{bh} frame and, thus, it probes the strong gravity region of the spacetime.  We consider the special case in which the \acs{bh} velocity is along its spin axis. Here the rate of change of the \acs{bh}'s energy is
\begin{eqnarray}
\dot{E}'_\text{BH}&=&\frac{16 \pi M^2n'\hbar \omega'(1+v)v}{1-v}\nn \\
&&\times \left\{ \frac{\ell_\text{cr}^2}{16}\left(\frac{1}{v}-1\right)-\log\left(\frac{b_\text{max}}{20M}\right)-\Lambda^2\right\},
\end{eqnarray}  
and the force acting on the moving \acs{bh} is
\begin{eqnarray}
\bm{F}'&\simeq&-\frac{16 \pi M^2n'\hbar \omega'(1+v)\bm{v}}{(1-v)v}\nn \\
&&\times \left\{ \log\left(\frac{b_\text{max}}{20M}\right)+\Lambda^2+ \frac{\ell_\text{cr}^2}{16}(1-v)\right\}.
\end{eqnarray}
These expressions are valid for media with~$b_\text{max}\geq20M$ and the function~$\Lambda(\tilde{a})$ is fitted by~\eqref{Lambda}. These analytical expressions describe excellently the numerical results of~\cite{Cardoso:2019dte}.

\newpage

\bibliography{References}

\end{document}